\documentclass[AMA,STIX1COL]{WileyNJD-v2}
\usepackage{moreverb}

\usepackage{float}
\usepackage{amsmath} 
\usepackage{bbm}
\usepackage{latexsym}
\usepackage{comment}
\usepackage{threeparttable} 
\usepackage{arydshln} 
\usepackage{subcaption}
\usepackage{graphics} 
\usepackage{hyperref}
\usepackage{bm} 
\usepackage{threeparttable} 

\usepackage{pifont} 
\usepackage{xr}
\makeatletter
\newcommand*{\addFileDependency}[1]{
\typeout{(#1)}
\@addtofilelist{#1}
\IfFileExists{#1}{}{\typeout{No file #1.}}
}\makeatother

\newcommand*{\myexternaldocument}[1]{%
\externaldocument{#1}%
\addFileDependency{#1.tex}%
\addFileDependency{#1.aux}%
}

\myexternaldocument{supplementary_material}

\newcommand\BibTeX{{\rmfamily B\kern-.05em \textsc{i\kern-.025em b}\kern-.08em
T\kern-.1667em\lower.7ex\hbox{E}\kern-.125emX}}

\articletype{RESEARCH ARTICLE}%

\received{\today}
\revised{<day> <Month>, <year>}
\accepted{<day> <Month>, <year>}

\begin{document}

\title{Model Selection for Causal Modeling in Missing Exposure Problems}

\author[1]{Yuliang Shi*}

\author[1]{Yeying Zhu}

\author[1]{Joel A. Dubin}

\authormark{Yuliang \textsc{et. al.}}

\address[1]{\orgdiv{Department of Statistics and Actuarial Science}, \orgname{University of Waterloo}, \orgaddress{\state{Waterloo}, \country{Canada}}}



\corres{*Yuliang Shi. \email{yuliang.shi@uwaterloo.ca}}

\presentaddress{200 University Ave W, Waterloo, ON, Canada, N2L 3G1.}

\abstract[Abstract]{In causal inference,  properly selecting the propensity score (PS) model is an important topic and has been widely investigated in observational studies. There is also a large literature focusing on the missing data problem. However, there are very few studies investigating the model selection issue for causal inference when the exposure is missing at random (MAR). In this paper, we discuss how to select both imputation and PS models, which can result in the smallest root mean squared error (RMSE) of the estimated causal effect in our simulation study. Then, we propose a new criterion, called  ``rank score'' for evaluating the overall performance of both models. The simulation studies show that the full imputation plus the outcome-related PS models lead to the smallest RMSE and the rank score can help select the best models.  An application study is conducted to quantify the causal effect of cardiovascular disease (CVD) on the mortality of COVID-19 patients.}


\maketitle

\section{Introduction}
\label{sec:intro}

In application studies, missing data can occur in multiple ways and one of the common cases is that the exposure of interest may not be fully observed \cite{rubin1976inference}. For example, a patient's exposure status may not be fully recorded when the patient suddenly drops out of the clinical study; or an individual may decline to answer a sensitive question regarding his/her health problem in a survey questionnaire. 
In addressing the challenge of missing data, researchers have proposed various methodologies for handling cases where the outcome is missing \cite{Bang}. However,  when the exposure status is missing, few approaches have been provided in literature to deal with this issue \cite{williamson2012doubly,zhang}. 

In causal inference,  propensity score (PS) analysis is a popular tool to address the confounding issue \cite{rosenbaum1983central}. However,  when the exposure status is missing at random (MAR), estimation of the causal effect is challenging. The case of MAR on exposure significantly differs from MAR on the outcome. When the outcome is missing, only the estimation of the outcome model is influenced by the missing data, but the estimation of the PS model will not be affected by the missing outcome. In contrast, when the exposure is missing, the estimation of all three models — imputation, PS, and outcome models— is affected. Consequently, although PS-based methods have been widely discussed when the outcome is MAR, their application to cases where the exposure is missing is not straightforward \cite{williamson2012doubly}. Careful adjustment regarding both missing and confounding issues are necessary in such scenarios.

One of the common approaches to dealing with missing data is via imputation. Single imputation (SI) is easy to conduct based on the observed data \cite{donders2006gentle}. However, imputing data only once may not be reliable and a large variation may be induced \cite{Buuren}.  In contrast, the multiple imputations chained equations (MICE) approach is proposed through random sampling, and  Rubin's Rules can be applied to account for both the sampling variability and the uncertainty in the imputation of missing values \cite{buuren2011,rubin2004multiple}.


In addition to the imputation-based method,  inverse probability weighting (IPW) or double robust (DR) methods have been widely investigated in the literature \cite{neyman1923application,scharfstein1999adjusting,robins1994estimation}. In recent literature, when the exposure is MAR, two-layer DR estimators have been proposed to deal with both missingness and confounding issues \cite{williamson2012doubly}. Later, a triple robust estimator (TR) was proposed to protect against misspecification of the missingness and imputation models \cite{zhang}. Besides that, nonparametric estimators with efficient bounds have been proposed using the efficient influence function \cite{kennedy2020efficient}. Compared with the imputation-based method, DR or TR approach requires more flexible conditions for the model specifications \cite{Seaman}, but very few papers discuss the model selection problem when the exposure is MAR. Despite the above-mentioned more sophisticated methods for estimating causal effects when the exposure is MAR, in this article, we focus on the more intuitive and straightforward approach: we impute the missing exposure values first via MICE and employ propensity score-based methods, such as  IPW or DR, to estimate causal effects based on the imputed datasets. Therefore, we face the model selection issue for both the imputation model and the PS model.

In the previous literature, researchers have shown that improper model selection for the PS model can result in higher bias and variance \citep{brookhart2006variable}. However,  little attention has been given to addressing how to select both imputation and PS models when the exposure is MAR. As a motivating example, we consider the case in which researchers aim to determine the causal relationship between the incidence of cardiovascular disease (CVD, as the exposure) and COVID-19 patients' mortality (as the outcome) in an observational study where CVD status is not completely observed. In this situation, we need to make adjustment for both missing CVD status and other confounding factors. However, which variables should be selected for the imputation and PS models is not clear because we do not know the true missing values of CVD.

Another problem from the motivating example is the so-called ``double-dipping issue'', which means if we impute the missing values of exposure with all exposure-related covariates on the first step and fit the PS model with the same set of covariates using that imputed dataset on the second step, the bias and variance of the estimated causal effect may be increased due to two mains reasons: (1) we do not clearly distinguish the purpose of fitting the imputation and PS models; (2) we include redundant exposure-related covariates instead of outcome-related variable into two models. More specifically, for the imputation model, we aim to increase its predictive ability to correctly impute the missing exposure. In contrast, to select the proper PS model, we intend to balance the confounders across exposure and non-exposure groups instead of increasing its predictive performance \cite{rubin2004principles}. In addition, including all exposure-related covariates in the PS model is also redundant because those variables are not considered as main confounders.  

Even though fitting the imputation and PS models with the same set of exposure-related covariates is very common in application studies, it is very important to investigate which variables should be included in the imputation and PS models to achieve the best performance of causal estimates among candidate models. 
Furthermore, since we generally do not know the true causal effect in the application, evaluating those selected models through some model selection criteria is also essential. Therefore,  we propose a new rank-based criterion to combine the performance of the imputation and PS models, compared with the traditional criterion. The rest of this article is organized as follows. In Section 2, we describe the goal of our research and discuss the proposed criteria for conducting model selection. In Section 3, we further design a simulation study to investigate model selection in both imputation and PS models. In Section 4, an application study is conducted to identify the causal effect of CVD on mortality for a cohort of COVID-19 patients. The article ends with a discussion in Section 5.

\section{Notation and Methods}
\label{sec:model}

\subsection{Framework and Assumptions}
\label{sec:framework}

Based on the counterfactual causal framework \cite{rosenbaum}, we denote $(Y_i^1, Y_i^0), i=1,\dots,n,$ as the potential outcomes if individual $i$ were exposed or unexposed (or equivalently,  treated or untreated), respectively.  Let $\bm{X}_i$ denote the covariates, $A_i$ denote the exposure of interest, $Y_i$ denote the binary outcome, and $R_i=I\{A_i \text{ is missing}\}$ denote the indicator of whether the exposure value is missing or not for individual $i=1,\dots,n$. In this study, we focus on the exposure variable being missing, so all other variables are assumed to be completely observed. In addition, we study the model selection problem on a finite set of covariates instead of the high-dimensional setting.

For each individual $i$, $Y_i^1$ and $Y_i^0$ cannot be observed at the same time; instead, we only observe $Y_i$ and $A_i$. The observed dataset is $(\bm{X}_i, A_i, Y_i, R_i), i=1, \dots, n$, and the relationship between the observed and potential outcomes can be written as $Y_i=A_iY_i^1+(1-A_i)Y_i^0$, for $i=1,\dots,n$. The true propensity score (PS) is defined as $\pi(\bm{X}_i)=P(A_i=1|\bm{X}_i)$ for individual $i=1,\dots,n$. We denote $\tau_1=E(Y^1)$ and $\tau_0=E(Y^0)$ as the average potential outcomes if treated or untreated, respectively. For the binary outcome in our application studies, we define the causal estimand of interest as the causal risk ratio: $\tau=\tau_1/\tau_0$.  Even though the study focuses on the binary outcome, the theory is developed in general scenarios and the key ideas presented in this article will not be largely changed by the different formats of the causal quantities, so the application can be easily expanded to other cases, such as the continuous outcome or count outcome. To estimate the causal risk ratio $\tau$, three assumptions are required as follows:
\begin{enumerate}

    \item Strongly ignorable treatment assignment assumption (SITA): $(Y^1, Y^0) \bot A|\bm{X}$. 
    
    \item Positivity assumption:  $0<P(A=1|\bm{X}=\bm{x})<1, \text{for all possible }  \bm{x}$.


    \item MAR assumption: $R \bot A|\bm{X},Y$.
\end{enumerate}


Here, SITA assumption is also known as the assumption of no unmeasured confounders, which cannot be tested in the application studies. The positivity assumption requires the propensity score to be a positive probability, which can be checked after imputing the missing exposure status correctly  \cite{rosenbaum}. MAR assumption assumes that the missing indicator is conditionally independent of the exposure itself given all observed covariates and outcomes in the dataset, which also cannot be tested on the observed data due to the unknown missing values \cite{williamson2012doubly}. 

The objective of this study is two-fold. Firstly, since different components of $\bm{X}$ can be different as shown in Figure \ref{fig:causal_sim}, our primary goal is to find the best selection on both imputation and PS models when the exposure is MAR. Secondly, we aim to find a suitable criterion to evaluate the performance of the two models in the application study when a DAG is not known.

\subsection{An Illustrative  Example}

\begin{figure}
\centering
\includegraphics[scale=0.65]{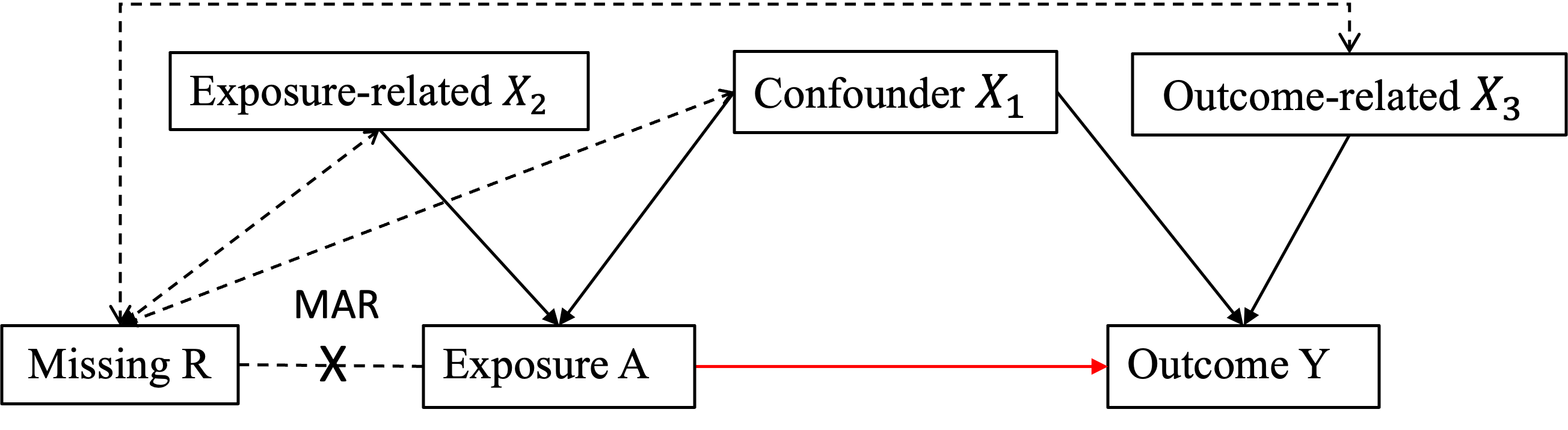}
\caption{An illustrative causal diagram for simulation studies. The black arrows refer to the causal relationship among confounders, the exposure, and the outcome. The double-sided dash arrows refer to the associational relationship among the missing indicator, the outcome, and the covariates. The red arrow is the causal effect of primary interest. The dashed short line refers to no association between the missing indicator and the exposure given covariates and the outcome due to MAR assumption.}
\label{fig:causal_sim}
\end{figure}

To study the different roles of covariates in model selection, we consider a simplified scenario with three covariates, which have different effects on the exposure or the outcome. As shown in Figure \ref{fig:causal_sim}, $X_1$ is the main confounder, $X_2$ is the exposure-related covariate, and $X_3$ is the covariate related only to the outcome. We denote the $4 \times 1$ vector of covariates for subject $i$ as $\bm{X}_i=(1,X_{i1},X_{i2},X_{i3})^T$. In terms of assumptions described in the general scenario in Section \ref{sec:framework}, we can rewrite three main assumptions based on Figure \ref{fig:causal_sim} as follows:
\begin{enumerate}

    \item Strongly ignorable treatment assignment assumption (SITA): $(Y^1, Y^0) \bot A|X_1$. 
    
    \item Positivity assumption:  $0<P(A=1|\bm{X}=\bm{x})<1, \text{for all possible }  \bm{x}$.


    \item MAR assumption: $R \bot A|X_1,X_2$.
\end{enumerate}

Since $X_1$ is the ``sufficient set'' for confounding adjustment in this specific case \cite{tanner2014identifying}, SITA assumption only requires conditioning on the main confounder $X_1$ instead of all covariates. In addition, we do not include the association between $R$ and $Y$ as shown in Figure \ref{fig:causal_sim} since we want to investigate whether adding $X_3$ and $Y$ will increase the accuracy of the imputation model for the missing exposure. In that way, we only need to condition on $X_1, X_2$ for MAR assumption based on Figure \ref{fig:causal_sim}.

To investigate the role of each variable in the imputation model, one of the approaches is to rely on the theories of directed acyclic graph (DAG), which is a common tool using ``d-separation'' \citep{glymour2016causal} to check conditional independence among variables without specifying the form of the models. In our specific example in  Figure \ref{fig:causal_sim}, we know that $X_1, X_2$ should be included in the imputation model because they affect $A$ and $R$ directly. The main question is whether $Y$ should be included in the imputation model or not. Even though we do not include the outcome in the true missingness model in Figure \ref{fig:causal_sim}, $Y$ is directly correlated with the exposure variable, so adding $Y$ may help predict the missing exposure values. 

Next, we aim to study the role of $X_3$ in the imputation model. Notice that  $X_3 \bot A$, but if given $(X_1,X_2)$, $Y$ is a function of $X_3$ and $R$ is also correlated with $X_3$, as shown in Figure \ref{fig:causal_sim}. 
In addition, if we condition on $Y$, $Y$ will be the collider in the path of $A \rightarrow Y \leftarrow X_3$, so $X_3 \not\!\perp A|(Y,X_1,X_2)$. In other words, including the outcome and outcome-related variables in the imputation model is expected to improve the predictive ability of the imputation model. In summary, we should include all $(X_1,X_2,X_3,Y)$ into the imputation model, which will be verified later in simulation studies.

\subsection{Estimation}

\label{sec:estimation}

Before we discuss the model selection strategy, we first discuss how to estimate $\tau$. As we have previously mentioned, estimating $\tau$ requires us to account for both missing and confounding issues. For the missing data problem, we will impute the exposure status via MICE based on MAR assumption \citep{Buuren,buuren2011}. After the missing exposures are imputed, $\tau$ can be estimated using inverse-probability weighting (IPW) or double robust (DR) method to account for the confounding issues.

In summary, we present an algorithm with three key steps as follows: 
 \begin{enumerate}
    \item Fit imputation (Imp) model based on MAR assumption using MICE: fit logistic regression model for $A \sim \bm{X}+Y$ to obtain $A_i^{\text{imp}}$ as the exposure on the imputed dataset, which includes the original exposure for individuals without missing values and imputed exposure for individuals with missing values. 
             
    \item Fit PS model using the logistic model on the imputed dataset: $A^{\text{imp}} \sim \bm{X}$ to obtain fitted PS values, denoted as $\hat{\pi}_i(\bm{X})$.
            
    \item Apply inverse weighting to adjust for the confounding issue,  and the IPW estimator can be written as \citep{Thompson}: 
\begin{equation}
\begin{aligned}
\hat{\tau}_1^{\text{IPW}}=\frac{1}{n} \sum_{i=1}^n \frac{A_i^{\text{imp}}Y_i}{\hat{\pi_i}(\bm{X}_i)}, \ \
\hat{\tau}_0^{\text{IPW}}=\frac{1}{n} \sum_{i=1}^n \frac{(1-A_i^{\text{imp}}) Y_i}{1-\hat{\pi_i}(\bm{X}_i)}
\end{aligned}
\end{equation}

Then, an IPW estimator for $\tau$ is: $\hat{\tau}^{\text{IPW}}=\hat{\tau}_1^{\text{IPW}}/\hat{\tau}_0^{\text{IPW}}$ or the DR estimator can be written as:
\begin{equation}
\begin{aligned}
\hat{\tau}_1^{\text{DR}}=\frac{1}{n} \sum_{i=1}^n \left[\frac{A_i^{\text{imp}}Y_i}{\hat{\pi_i}(\bm{X}_i)}- \frac{\hat{\pi_i}(X)-A_i^{\text{imp}} }{\hat{\pi_i}(\bm{X}_i)} \hat{m_1}(A_i^{\text{imp}},\bm{X}_i) \right]
,  \ \
\hat{\tau}_0^{\text{DR}}=\frac{1}{n} \sum_{i=1}^n \left[ \frac{(1-A_i^{\text{imp}})Y_i}{1-\hat{\pi_i}(\bm{X}_i)}- \frac{A_i^{\text{imp}}-\hat{\pi_i}(X) }{1-\hat{\pi_i}(\bm{X}_i)} \hat{m_0}(A_i^{\text{imp}},\bm{X}_i)\right] 
\end{aligned}
\end{equation}
where $\hat{m_1}(A_i^{\text{imp}},\bm{X}_i)$ is the fitted response for the treatment group, i.e., $\hat{m_1}(A_i^{\text{imp}},\bm{X}_i)=E[Y_i|A_i^{\text{imp}}=1,\bm{X}_i;\hat{\bm{\beta}}_1]$; $\hat{m_0}(A_i^{\text{imp}},\bm{X}_i)$ is the fitted response for the control group, i.e., $\hat{m_0}(A_i^{\text{imp}},\bm{X}_i)=E[Y_i|A_i^{\text{imp}}=0,\bm{X}_i;\hat{\bm{\beta}}_0]$. Here, $(\hat{\bm{\beta}_1},\hat{\bm{\beta}_0})$ are estimated parameters for the treatment or control group from the outcome model, written as $Y \sim A+X$.  Then, a DR estimator for $\tau$ is: $\hat{\tau}^{\text{DR}}=\hat{\tau}_1^{\text{DR}}/\hat{\tau}_0^{\text{DR}}$.

\end{enumerate}


Notice that when we conduct imputation using MICE,  we choose $m=20$  as the number of imputations. Then, IPW and DR estimators for $\tau$ can be constructed on each imputed dataset and Rubin's Rules is applied to obtain final estimated values based on the algorithm described above \ref{sec:estimation} \cite{rubin2004multiple}. If the imputation model is correctly specified, based on MAR assumption,  we can approximate  $P(A=1|\bm{X},Y)$ by $P(A^{\text{imp}}=1|\bm{X},Y)$. Furthermore, if PS model is also correct, due to SITA assumption, $\hat{\tau}^{\text{IPW}} \xrightarrow[]{\text{p}} \tau$ and its asymptotic normality holds as $n \to \infty$ \cite{Thompson}. For DR estimator, it can protect against misspecification of either PS or the outcome model \cite{robins1994estimation,scharfstein1999adjusting}. In other words, if the imputation model is correct and either PS or the outcome model is correct, we know $\hat{\tau}^{\text{DR}} \xrightarrow[]{\text{p}} \tau$ and its asymptotic normality holds as $n \to \infty$ \cite{Bang}. 


\subsection{Model Selection Criteria}
\label{sec:model_criteria}

In application studies, since the true causal effect is unknown, we are not able to find which combination of imputation and PS models leads to the best performance in terms of some performance metrics, such as RMSE. In such a case, we need some model selection criteria to choose the best combination of imputation and PS models.  In this section, we discuss some traditional criteria for model selection and propose a new criterion that takes into consideration both models. 

\subsubsection{Weighted Accuracy ($\text{Accuracy}^{(w)})$}
\label{sec:weighted_acc}

We first discuss a theoretical way to evaluate the performance of the imputation model via weighted accuracy based on the observed data. Due to the missing exposure, even though we can impute the missing data, we do not know the true missing values, which makes it challenging to estimate the accuracy only based on the observed data.  A naive approach is to fit the imputation model based on the observed data and compare the imputed exposure with those observed values. However, since the accuracy is a function of $(\bm{X},Y)$ and the distribution of $(\bm{X},Y)$ on observed data is different from the whole data in general, this approach to estimate accuracy from the observed data is no longer valid.


One approach to deal with this issue is to apply inverse probability weighting on the accuracy when we consider $w_i=P(R_i=1|\bm{X_i},Y_i)$ as the propensity for the missingness. Obtaining the weighted accuracy can be described in the following four steps:
\begin{enumerate}
    \item 
    First, we randomly split the individuals into either the training or the testing data according to a ratio $q$, where $q=\frac{n_{\text{test}}}{n_{\text{obs}}}$. Here, 
 $n_{\text{obs}},n_{\text{test}}$ is the sample size for the whole observed data and for the testing data, respectively. Certainly, $q$ can be arbitrarily chosen by the user, but the simplest way is to set the ratio equal to the missing rate in the original dataset.

    \item Next, we fit the imputation model on the training data and impute the exposure on the testing data to compare with the known exposure status.
    
    \item  Then, we fit the full missingness model on the whole dataset, written as $\text{logit}(w_i)=\bm{X_i^T \gamma}$, where $\bm{\gamma}$ is the vector of coefficients including the intercept to be estimated so that we can obtain $\hat{w_i}$ as the estimated propensity of missingness for individual $i=1,2,\dots,n$.

    \item Finally, we calculate the weighted accuracy based on the observed data, called ``$\text{Accuracy}^{(w)}$'':
\begin{equation}
\label{est:ipw_acc}
\begin{aligned}
\text{Accuracy}^{(w)}(A^{\text{imp}})= \frac{\sum_{i=1}^n \frac{1-R_i}{1-\hat{w}_i} \delta_i \mathbbm{1}(A_i=A^{\text{imp}}) }{n\times q},
\end{aligned}
\end{equation}
\end{enumerate}
where $\delta_i=\mathbbm{1}(\text{individual $i$ is chosen in the test set})$ and $\mathbbm{1}(A_i=A^{\text{imp}})$ is an indicator for whether the imputed value equals to the observed value. Here, $\hat{w}_i$ is estimated from the full missingness model including all covariates and the outcome. The weighted accuracy is consistent to the true accuracy when the full missingness model is correctly specified and the detailed proof is attached in Appendix \ref{app:proof_acc}.

\subsubsection{ASMD, KS, and BIC}
\label{sec:guideline}

To evaluate the PS model, the traditional approaches are based on the balance statistic calculated from the confounders. For example,  one can calculate absolute standardized mean difference (ASMD) and Kolmogorov-Smirnov (KS)  statistics in the covariates after propensity score adjustment \cite{lilliefors1967kolmogorov,berger2014kolmogorov,franklin2014metrics}. 

Another approach to evaluate the PS model is using BIC criterion to select the best model adjusting for both the goodness of fit and the number of parameters in the model. Since in PS model selection, it is recommended to select the outcome-related variables instead of the exposure-related variables \cite{brookhart2006variable,bhattacharya2007instrumental,zhu2015variable}, we suggest using BIC of the outcome model, i.e.,  $Y \sim A+\bm{X}$, as the selection criterion. For example, from Figure \ref{fig:causal_sim}, since $X_2 \bot Y|A=a$, a smaller BIC indicates that we have included outcome-related variables and excluded the exposure-related variables. Notice that we do not recommend using the c-statistics or accuracy of PS model to select variables because in PS analysis, we aim to balance the confounders across the exposure and non-exposure group instead of maximizing the predictability of the PS model \cite{rubin2004principles,westreich2011role,patrick2011implications}. 


We should also notice that only using one traditional criterion may not be appropriate to select both imputation and PS models, such as either $\text{Accuracy}^{(w)}$, ASMD, or KS statistics, because they just focus on the performance of a single model. That motivates us to combine the evaluation of $\text{Accuracy}^{(w)}$ and BIC into an integrated criterion, called ``rank score'', proposed in the next section. 

\subsubsection{Rank Score}
\label{sec:rank_score}

The idea of this new criterion called ``rank score'' is to take into account of the performance of both imputation and PS models. We first obtain the values of $\text{Accuracy}^{(w)}$ and BIC for all possible combinations of the candidate models.  Then, for a given model, we calculate its rank score value by:

\begin{equation}
\begin{aligned}
\text{ Rank Score}= \frac{\text{Rank(1-$\text{Accuracy}^{(w)}$)+Rank(BIC)}}{2},
\end{aligned}
\end{equation}
where ``Rank($1-\text{Accuracy}^{(w)}$)'' is the rank of the model based on the value of  $1-\text{Accuracy}^{(w)}$ if we order $1-\text{Accuracy}^{(w)}$ from the largest to the smallest. ``Rank(BIC)'' is the rank for a given model based on the value of BIC for regressing $Y$ on $A$ and the covariates selected in the given model.  We can calculate the rank score for every possible combination of the imputation and PS models. Then, we select the smallest rank score, which leads to the highest $\text{Accuracy}^{(w)}$ and the smallest BIC, so the best imputation and PS models can be successfully chosen. 

In summary, the main advantage of the rank score is to combine the performance of both imputation and PS models and directly find the best model based on this rank-based criterion. In addition, the rank score is a unit-free score, which is not affected by the different magnitudes of the accuracy and BIC.  The evaluation of those criteria is shown in Table \ref{table:sim_cor_n500}. 




\subsubsection{Other Criteria}



Another possible criterion to combine accuracy and BIC is to re-scale both terms into the range of  $[0, 1]$. Then, we average two rescaled terms and call the following criterion as ``ABIC'':
\begin{equation}
\begin{aligned}
\text{ABIC}= \frac{1}{2} \left[ \left(1-\frac{\text{$\text{Accuracy}^{(w)}$}-\text{min}(\text{$\text{Accuracy}^{(w)}$})}{\text{max(\text{$\text{Accuracy}^{(w)}$})}-\text{min(\text{$\text{Accuracy}^{(w)}$})}} \right)+\frac{\text{BIC}-\text{min}(\text{BIC})}{\text{max(\text{BIC})}-\text{min(\text{BIC})}} \right] ,
\end{aligned}
\end{equation}
where ``min'' means the smallest value and ``max'' means the largest value among all candidates of models. Notice that we cannot directly average over the accuracy and BIC values because of the different magnitude issues. Therefore, ABIC is considered to solve that problem after we rescale the accuracy and BIC between 0 and 1. We want to choose the candidate model with a smaller ABIC value, which usually means that its accuracy will be larger and BIC will become smaller.  However, the min and max values of ABIC can still be affected by the extreme values of either BIC or $\text{Accuracy}^{(w)}$. As a result, the  ABIC values among the candidate models may be largely shrunk to a quite small value, which makes it hard for us to distinguish the model performance. The main results are shown in Table \ref{table:sim_ipw_n500_acc} and \ref{table:sim_dr_n500_acc} in Section \ref{sec_simdr_binary} of Appendix.






\section{The Simulation Studies}
\label{sec:sim}

\subsection{Simulation Setup}

To investigate which covariates should be included in the imputation and PS models, a simulation study is conducted based on the causal diagram in Figure \ref{fig:causal_sim}. The number of replications is $N=1000$,  and the sample size is $n=500$ in each data generation. Three covariates $X_1,X_2,X_3$ are generated from $N(0,1)$ independently.  $Y, A$, and the missing indicator $R$ are generated by the following three models:
\begin{itemize}
    \item missingness model: $\text{logit}\{P(R=1|\bm{X})\}=-0.3+0.4X_1+0.6X_2+1.8X_3$;
    \item treatment model: $\text{logit}\{P(A=1|\bm{X})\}=-0.2+0.3X_1+1X_2$;
    \item outcome model: $\text{logit}\{P(Y=1|A,\bm{X})\}=-0.2+2A-0.3X_1+2.5X_3;$
\end{itemize}

The missing rate is about 48\% in this scenario and the true causal effect $\tau=1.523$. Notice that the true missingness model does not include the outcome $Y$, but one of our aims is to investigate whether adding the outcome into the imputation model will improve the performance of the estimated causal effect in the simulation studies. 

For each simulated data set, we investigate five different imputation models and four different PS models, which we have specified and named in Tables \ref{table:imp_modelset}, respectively.  In total, twenty possible combinations of the imputation and PS models are investigated. We provide some rationale behind the specification of these imputation and PS models. When selecting the imputation model, we start with an exposure-related model, which includes only $X_1$ and $X_2$ as they directly affect $A$. To increase the predictive ability, we gradually modify the imputation model by adding $X_3$ and $Y$ in a step-by-step process until the full imputation model is obtained. 

\begin{table}
    \caption{Imputation and PS models in the simulation studies}
    \label{table:imp_modelset}
    \centering
    \begin{tabular}{cc}
    \hline
       Model Name (shortened form) & Model Form  \\
    \hline
   \multicolumn{2}{c}{Five imputation models} \\
     exposure-related model (Exp)  &  $A \sim X_1+X_2$ \\
    outcome-related model (Out)  & $A \sim X_1+X_3$  \\
    all covariates-included model (Covs) & $A \sim X_1+X_2+X_3$   \\    
    response-included model (Res) &$A \sim X_1+X_2+Y$  \\
    full model (Full) & $A \sim X_1+X_2+X_3+Y$ \\
          \hdashline
    \multicolumn{2}{c}{Four PS models}  \\
     naive model (Naive)  &  $A^{\text{imp}} \sim X_1$  \\
    exposure-relate model (Exp)  & $A^{\text{imp}} \sim X_1+X_2$  \\
   outcome-relate model (Out)  & $A^{\text{imp}} \sim X_1+X_3$  \\
    all covariates-included model (Covs) & $A^{\text{imp}} \sim X_1+X_2+X_3$  \\
          \hline
    \end{tabular}
\end{table}

In contrast, the goal of fitting PS model is not to increase the predictive ability, but to adjust for the confounding issue. Therefore, we start with a naive PS model with only $X_1$ as the main confounder. Then, we gradually add exposure-related or outcome-related variables until PS model includes all covariates.

\subsection{Simulation Results for Imputation and PS Selection}

\begin{sidewaysfigure}
\centering
\includegraphics[width=\textwidth,height=10cm]{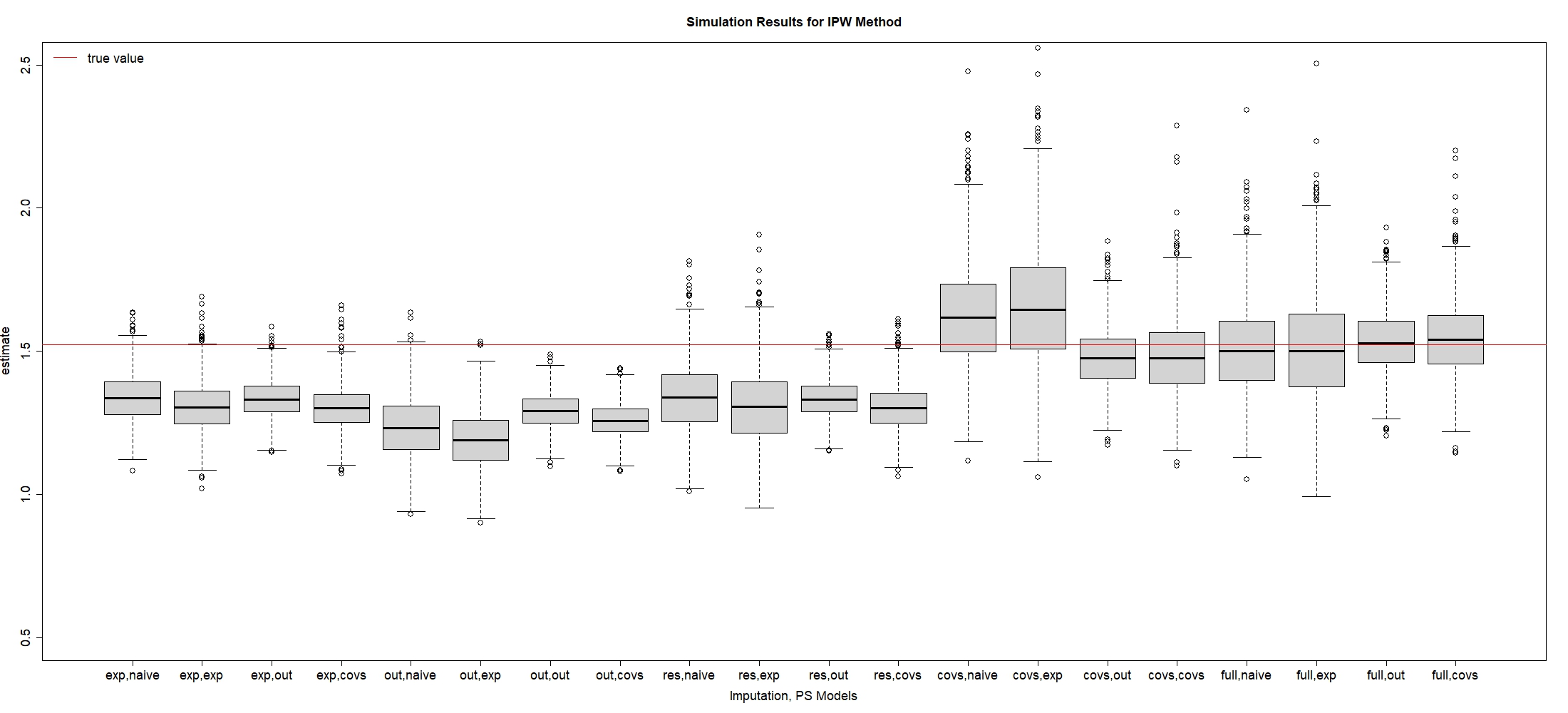}
\caption{Boxplot of estimated values for IPW estimators in the simulation studies when $n=500$. The x-axis refers to the combination of imputation and PS models. Here, Exp, Out, Res, and Covs, refer to the exposure-related, outcome-related, response-included, and covariates-included models, respectively. The red line is the true causal effect. } 
\label{figure:sim_ipw_n500}
\end{sidewaysfigure}

In this subsection, we investigate which combination of imputation and PS models can lead to the optimal RMSE of $\hat{\tau}$. Table \ref{table:sim_ipw_n500} shows simulation results for IPW estimator when $n=500$, and Figure \ref{figure:sim_ipw_n500} is plotted to visualize these results. We also provide the simulation results for the continuous outcome with the goal of estimating the average causal effect in Section S.1 of the supplementary material. To evaluate the performance of different models, we present the bias rate as the quantity of main interest: $(\Bar{{\tau}}-\tau)/\tau$, where $\Bar{{\tau}}=\frac{1}{N} \sum_{j=1}^N \hat{{\tau}_j}$, and $\hat{{\tau}}_j$ is the point estimate in the $j^{th}$ replication, $j=1,2,\dots, N$. The empirical standard error (ESE), which is also called the sampling standard error, is written as $\text{ESE}=\sqrt{\frac{1}{N-1} \sum_{j=1}^N(\hat{{\tau}_j}-\Bar{{\tau}})^2}$. The square root of mean squared error (RMSE) is defined as $\text{RMSE}=\sqrt{\frac{1}{N-1}\sum_{j=1}^N(\hat{{\tau}}_j-\tau)^2}$.

Since $(X_1,X_2)$ are both directly correlated with the exposure from Figure \ref{fig:causal_sim}, some researchers may argue that we only need to include $(X_1,X_2)$ without $(X_3,Y)$ into the imputation and PS models as the best results. However, Table \ref{table:sim_ipw_n500} shows that the exposure-related imputation model plus the exposure-related PS model leads to higher bias and variance than the full imputation model or the outcome-related PS model. That provides support to compare the performance among different candidates of imputation and PS models in the following sections.

\begin{table}
\caption{Performance of different models using IPW method when $n=500$} 
\label{table:sim_ipw_n500}
\centering
\begin{tabular}{lcccc}
 \hline
Imp, PS Models & Bias & Bias Rate & ESE  & RMSE \\ 
\hline
Exp, naive & -0.186 & -12.187 & 0.085 & 0.204 \\ 
  Exp, exp & -0.217 & -14.283 & 0.092 & 0.236 \\ 
  Exp, out & -0.188 & -12.324 & 0.067 & 0.199 \\ 
  Exp, covs & -0.219 & -14.384 & 0.080 & 0.233 \\ 
    \hdashline
  Out, naive & -0.290 & -19.069 & 0.109 & 0.310 \\ 
  Out, exp & -0.333 & -21.858 & 0.103 & 0.348 \\ 
  Out, out & -0.230 & -15.074 & 0.059 & 0.237 \\ 
  Out, covs & -0.263 & -17.301 & 0.059 & 0.270 \\ 
    \hdashline
  Covs, naive & -0.181 & -11.920 & 0.123 & 0.219 \\ 
  Covs, exp & -0.211 & -13.883 & 0.138 & 0.252 \\ 
  Covs, out & -0.188 & -12.335 & 0.067 & 0.199 \\ 
  Covs, covs & -0.218 & -14.295 & 0.080 & 0.232 \\ 
    \hdashline
  Res, naive & 0.102 & 6.679 & 0.176 & 0.204 \\ 
  Res, exp & 0.137 & 9.026 & 0.217 & 0.257 \\ 
  Res, out & -0.046 & -2.996 & 0.107 & 0.116 \\ 
  Res, covs & -0.038 & -2.523 & 0.137 & 0.143 \\ 
    \hdashline
  Full, naive & -0.017 & -1.096 & 0.160 & 0.161 \\ 
  Full, exp & -0.011 & -0.710 & 0.194 & 0.194 \\ 
  Full, out & 0.010 & 0.640 & 0.110 & \textbf{0.111} \\ 
  Full, covs & 0.023 & 1.498 & 0.137 & 0.139 \\ 
   \hline
\end{tabular}
 \begin{tablenotes}
      \item  
 Here, Exp, Out, Res, and Covs, refer to the exposure-related, outcome-related, response-included, and covariates-included models, respectively. The lowest RMSE is bolded above. 
    \end{tablenotes}
\end{table}

\subsubsection{Comparison of Imputation Models}

Based on Table \ref{table:sim_ipw_n500} and Figure \ref{figure:sim_ipw_n500}, if we focus on the same PS model and compare different imputation models, we find that the full imputation model has resulted in the smallest bias of $\hat{\tau}$ compared to other imputation models. In contrast, the exposure-related, outcome-related, and all covariates-included imputation models generate huge bias rates.

Furthermore, our simulation results indicate that adding $(X_3,Y)$ into the imputation model is necessary to reduce the bias of the estimated causal effect. For example, comparing the exposure-related model with the response-included imputation model, Table \ref{table:sim_ipw_n500} shows that adding $Y$ into the imputation model reduces about 10\% of bias. To study the effect of $X_3$, if we compare the exposure-related model versus the covariate-included imputation model, adding $X_3$ into the imputation model does not reduce the large bias if we do not condition on $Y$. In contrast,   when we condition on $Y$, compared with the response-included imputation versus the full imputation model, Figure \ref{figure:sim_ipw_n500} shows that adding $X_3$ leads to a reduction of bias. These findings are consistent with the discussion of DAG in Section \ref{sec:model}, which indicate that even though the true missing model does not include $Y$ and $A$ is not directly correlated with $X_3$, adding both $X_3$ and $Y$ into the imputation model can improve the predictive ability. 

In summary, even though the true missingness model does not include $Y$,  all exposure-related, outcome-related covariates and the outcome itself should be included in the imputation model, as the full imputation model results in the smallest bias among others. 


\subsubsection{Comparison of PS Models}

Based on the above discussion, we next focus on the full imputation model and compare the performance of different PS models in Table \ref{table:sim_ipw_n500} and Figure \ref{figure:sim_ipw_n500}. We find that all PS models that incorporate the main confounder $X_1$ can lead to very small bias as we have previously discussed that $X_1$ is the ``sufficient set'' in this case \cite{tanner2014identifying}. Even though $X_2$ directly causes $A$, adding $X_2$ does not reduce the bias because it only causes the change of the exposure and has no direct impact on the outcome $Y$, so $X_2$ is not considered as the main confounder. 

In terms of standard errors, the outcome-related PS model with $X_3$ can generate the smallest ESE when we fit the full imputation model. In contrast, both exposure-related and all-covariates PS models lead to an increase in ESE because adding exposure-related variable $X_2$ is redundant, which may generate more extreme PS values to enlarge the variation in finite samples \cite{brookhart2006variable}. Therefore, we recommend choosing the outcome-related PS model over the exposure-related PS model. 

The results using DR estimator are presented in Table \ref{table:sim_dr_n500} and visualized in Figure \ref{figure:sim_dr_n500} in Section \ref{sec_simdr_binary} of Appendix. Overall, we observe a similar trend of model selection using DR estimator as IPW estimator, suggesting that full imputation plus outcome-based PS models can effectively reduce RMSE of $\hat{\tau}$ compared to other candidate models.  


In conclusion, the best model selection is to choose the full imputation plus outcome-related PS models for both IPW or DR estimator, which results in the smallest RMSE of $\hat{\tau}$. In contrast, using outcome-related imputation plus exposure-related PS models leads to the largest RMSE, thus is not recommended.

\subsection{Simulation Results of Selection Criteria}
\label{sec:sim_criterion}

To evaluate the performance of different criteria discussed in Section \ref{sec:model_criteria}, we calculate the correlation between RMSE and the value of each criterion in Table \ref{table:sim_cor_n500}. More specifically, after $N=1000$ replications, we have obtained RMSE for all candidate models. Within each replication of simulation studies, we can evaluate each candidate model with different criteria. Then, in the $j^{\text{th}}$ replication, we can calculate the Spearman correlation between the proposed criterion and RMSE of $\hat{\tau}$. Table \ref{table:sim_cor_n500} shows the average of the Spearman correlations over 1000 replications.

\begin{table}
\caption{Criterion values for different models using IPW method when $n=500$} 
\label{table:sim_ipw_n500_acc}
\centering
\begin{tabular}{lcccccccc}
   \hline
Imp, PS Models & $\text{Accuracy}^{(w)}$ & Out BIC & ASMD & KS & ABIC & Rank Score & Rank(RMSE) \\ 
 \hline
Exp, naive & 0.586 & 677.561 & 0.161 & 0.130 & 0.608 & 11.789 & 10 \\ 
  Exp, exp & 0.586 & 720.437 & 0.022 & 0.130 & 0.675 & 14.272 & 14 \\ 
  Exp, out & 0.586 & 430.074 & 0.149 & 0.130 & 0.225 & 7.849 & 7 \\ 
  Exp, covs & 0.586 & 433.966 & 0.014 & 0.130 & 0.231 & 8.959 & 13 \\ 
    \hdashline
  Out, naive & 0.507 & 683.862 & 0.103 & 0.148 & 0.930 & 15.947 & 19 \\ 
  Out, exp & 0.507 & 724.226 & 0.031 & 0.148 &  0.993 & 18.306 & 20 \\ 
  Out, out & 0.507 & 434.737 & 0.075 & 0.148 & 0.545 & 12.460 & 15 \\ 
  Out, covs & 0.507 & 435.659 & 0.008 & 0.148 & 0.546 & 12.918 & 18 \\ 
    \hdashline
  Covs, naive & 0.585 & 676.668 & 0.170 & 0.136 & 0.612 & 11.875 & 11 \\ 
  Covs, exp & 0.585 & 719.239 & 0.033 & 0.136 & 0.678 & 14.354 & 16 \\ 
  Covs, out & 0.585 & 430.103 & 0.148 & 0.136 & 0.231 & 8.043 & 8 \\ 
  Covs, covs & 0.585 & 433.958 & 0.014 & 0.136 & 0.237 & 9.136 & 12 \\ 
    \hdashline
  Res, naive & 0.609 & 652.471 & 0.182 & 0.124 & 0.484 & 8.103 & 9 \\ 
  Res, exp & 0.609 & 694.546 & 0.040 & 0.124 & 0.549 & 10.450 & 17 \\ 
  Res, out & 0.609 & 411.095 & 0.154 & 0.124 & 0.110 & 3.993 & 2 \\ 
  Res, covs & 0.609 & 415.906 & 0.015 & 0.124 & 0.118 & 4.580 & 4 \\ 
    \hdashline
  Full, naive & 0.622 & 662.369 & 0.172 & 0.138 & 0.430 & 7.463 & 5 \\ 
  Full, exp & 0.622 & 705.051 & 0.034 & 0.138 & 0.497 & 9.941 & 6 \\ 
  Full, out & 0.622 & 401.525 & 0.149 & 0.138 & 0.027 & \textbf{1.989} & \textbf{1} \\ 
  Full, covs & 0.622 & 406.367 & 0.014 & 0.138 &  0.034 & 2.574 & 3 \\ 
   \hline
\end{tabular}
 \begin{tablenotes}
      \item  
   $\text{Accuracy}^{(w)}$: weighted accuracy on the observed data; Out BIC: BIC of the outcome model; ASMD: absolute standardized mean difference for $(X_1,X_2,X_3)$; KS: Kolmogorov-Smirnov distance for $(X_1,X_2,X_3)$; ABIC: rescaled product of Accuracy and BIC. Rank Score: average ranks of $(1-\text{Accuracy}^{(w)})$ and BIC. The lowest rank and RMSE are bolded above. 
    \end{tablenotes}
\end{table}

From Table \ref{table:sim_cor_n500}, we find that adding more variables to the imputation model increases the accuracy values, so the full imputation model is preferred over other candidate models. In addition, if we fit the same imputation model, the naive and exposure-related PS models generally have larger BIC than outcome-related and covariates-included PS models. However, using a single criterion to select both models leads to some limitations. For example, if we just adopt $\text{Accuracy}^{(w)}$ to select imputation and PS models, we cannot distinct between different PS models because the same imputation model will lead to the same $\text{Accuracy}^{(w)}$ value. Similarly, if we only employ BIC to select two models and we choose the same exposure-related PS model, we cannot easily distinguish the performance of the exposure-related vs covariates-included imputation models.

Compared with other criteria, the rank score shows a stronger correlation (close to 0.85) with the RMSE of $\hat{\tau}$ using either IPW method, which indicates that a smaller value of the rank score is more likely to result in a smaller RMSE. In other words, even though we do not know the true causal effect and RMSE in the application, the rank score is valid for researchers to evaluate the performance of imputation and PS models together. In addition, from Table \ref{table:sim_ipw_n500_acc} using IPW method and Table \ref{table:sim_dr_n500_acc} using DR method,  we can confirm that full imputation plus outcome-related PS models have a higher chance to be selected via the ``rank score'' criterion.

In summary, the full imputation plus outcome-related PS models will result in the smallest rank score as the best choice, consistent with the smallest RMSE. Our proposed rank score combines the performance of both imputation and PS models. As a rank-based criterion, the rank score shows a stronger correlation with the RMSE of $\hat{\tau}$ compared with other traditional criteria. Based on our simulation results, the rank score can also successfully select the full imputation plus the outcome-related PS models with the smallest RMSE as the best choice. In comparison, if we choose the outcome-related imputation plus exposure-related model, the rank score has the largest value, which is also consistent with the highest RMSE. 

\begin{table}[ht]
\centering
\caption{Spearman correlation between criteria and RMSE when $n=500$} 
\label{table:sim_cor_n500}
\begin{tabular}{lcc}
 \hline
Criterion & IPW Method & DR Method \\
  \hline
1-$\text{Accuracy}^{(w)}$  & 0.678 & 0.655 \\ 
  ASMD & -0.306 & -0.337 \\ 
  KS & 0.212 & 0.191 \\ 
  Out BIC & 0.682 & 0.702 \\ 
    \hdashline
  ABIC & 0.793 & 0.785 \\ 
  Rank Score & 0.841 & 0.837 \\ 
   \hline
\end{tabular}
\begin{tablenotes}
\item $\text{Accuracy}^{(w)}$: weighted accuracy on the observed data; Out BIC: BIC of the outcome model; ASMD: absolute standardized mean difference for $(X_1,X_2,X_3)$; KS: Kolmogorov-Smirnov distance for $(X_1,X_2,X_3)$; ABIC: rescaled product of Accuracy and BIC. Rank Score: average ranks of $(1-\text{Accuracy}^{(w)})$ and BIC. 
\end{tablenotes}

\end{table}

\section{Application Study}
\label{sec:apply}

\subsection{Background}
\label{sec:apply_intro}

COVID-19 is a highly contagious disease that can lead to severe clinical manifestations and long-term complications for patients. Some researchers have shown that the presence of cardiovascular disease (CVD) is associated with significantly worse mortality in COVID-19 patients \cite{kassir2020risk}. However, the associational analysis cannot answer the question of whether CVD directly causes higher mortality or whether other risk factors that lead to CVD are also a common cause of COVID-19 death. In this section, we present an application study on a COVID-19 dataset to investigate the causal effect of CVD (as the exposure) on the mortality of COVID-19 patients (as the outcome) when CVD status is not fully observed. We aim to select both imputation and PS models using the proposed methods and compare their performance. 

The data were collected during the second wave of the pandemic in Brazil from June 25th to December 31st, 2020. The dataset contains 2878 COVID-19 patients, of whom 1253 died from COVID-19. Although all patients were exposed to COVID-19, 548 patients had missing information on CVD, resulting in a missing rate of 19\%. Demographic and clinical variables were also collected including age, sex, and diabetes without missing data. A summary of the dataset is provided in Tables \ref{table:summaryapply} and\ref{table:sum_apply_miss}, stratified by the mortality or the missingness, respectively. The independence between mortality and covariates was tested using the Fisher test for categorical variables and the Mann-Whitney test for continuous variables. From Table \ref{table:summaryapply}, we find a significant association between age and mortality, which is a well-known risk factor. There are also significant associations between CVD and mortality, which may be influenced by both confounders and missingness in the exposure.  However, no significant association is identified between either diabetes or sex with mortality.

\begin{table}
\centering
\caption{Summary table of COVID-19 data stratified by mortality}
\label{table:summaryapply}
\begin{tabular}{lcccc}
   \hline
Covariate & Full Sample (n=2878) & Alive (n=1625) & Died (n=1253) & p-value \\ 
 \hline
\textbf{CVD} &  &  &  & \textbf{0.022} \\ 
  ~~~No & 802 (34) & 478 (36) & 324 (32) &  \\ 
  ~~~Yes & 1528 (66) & 835 (64) & 693 (68) &  \\ 
  ~~~Missing & \textbf{548} & \textbf{312} & \textbf{236} &  \\ 
    \hdashline
  \textbf{age} &  &  &  & \textbf{$<$0.001} \\ 
  ~~~Mean (sd) & 68.9 (16.3) & 65.4 (16.6) & 73.3 (14.7) &  \\ 
  ~~~Median (Min,Max) & 72 (0,107) & 67 (0,104) & 76 (0,107) &  \\ 
    \hdashline
  \textbf{sex} &  &  &  & 0.054 \\ 
  ~~~Male & 1551 (54) & 850 (52) & 701 (56) &  \\ 
  ~~~Female & 1327 (46) & 775 (48) & 552 (44) &  \\ 
    \hdashline
  \textbf{diabetes} &  &  &  & 0.3 \\ 
  ~~~No & 1117 (39) & 617 (38) & 500 (40) &  \\ 
  ~~~Yes & 1761 (61) & 1008 (62) & 753 (60) &  \\ 
   \hline
\end{tabular}
 \footnotetext{
      CVD: cardiovascular disease. For the continuous variable, it shows the mean, se, min, max, and median. For the categorical variable, its frequency and percentage of total samples are provided in the bracket. 
    }
\end{table}

\begin{table}
\centering
\caption{Summary table of COVID-19 data stratified by the missing indicator}
\label{table:sum_apply_miss}
\begin{tabular}{lcccc}
  \hline
Covariate & Full Sample (n=2878) & Observed (n=2330) & Missing (n=548) & p-value \\ 
  \hline
  \textbf{age} &  &  &  & 0.099 \\ 
  ~~~Mean (sd) & 68.9 (16.3) & 69 (16.6) & 68.3 (14.8) &  \\ 
  ~~~Median (Min,Max) & 72 (0,107) & 72 (0,104) & 71 (1,107) &  \\ 
    \hdashline
  \textbf{sex} &  &  &  & 0.37 \\ 
  ~~~Male & 1551 (54) & 1246 (53) & 305 (56) &  \\ 
  ~~~Female & 1327 (46) & 1084 (47) & 243 (44) &  \\ 
    \hdashline
  \textbf{diabetes} &  &  &  & \textbf{$<$0.001} \\ 
  ~~~No & 1117 (39) & 1111 (48) & 6 (1) &  \\ 
  ~~~Yes & 1761 (61) & 1219 (52) & 542 (99) &  \\ 
   \hline
\end{tabular}
\end{table}

In this application, we aim to estimate the causal risk ratio of mortality for patients with CVD versus those without CVD as our main interest in the application study. However, in such observational studies, we cannot simply randomize the exposure status for patients, making it challenging to estimate the causal effect of CVD on COVID-19 mortality in the presence of confounding issues. Meanwhile, since CVD status is not fully observed, we also have to deal with the missing exposure problem, so we need to make assumptions about the missing exposure and confounding issues. From Table \ref{table:sum_apply_miss}, we find a significant association between missingness and diabetes. Older COVID-19 patients with chronic diseases, such as diabetes, may be more likely to report whether they have CVD or not, so assuming MCAR in this study is not reasonable. Instead, we adopt the MAR assumption, which assumes that the missing indicator for CVD is conditionally independent of CVD status itself, given other observed variables. In addition, we assume SITA assumption holds, i.e. the potential outcomes are conditionally independent of the CVD given all pre-measured covariates. After that, we intend to apply MICE to impute the missing values and IPW-based estimators to account for confounding issues.

\subsection{Model Selection}
\label{sec:apply_select}


Since the effect of each clinical variable on CVD status or the outcome may be different, properly selecting imputation and PS models is essential to obtain a valid estimation of the causal effect. However, a DAG in this case is not readily available if we lack sufficient background information in the application. Previous clinical papers show that age and sex are important risk factors for death \cite{dana2020insight,abdi2020diabetes}. In addition, many clinical studies find that diabetes and age can affect the risk of CVD \cite{bertoluci2017cardiovascular}. Based on the prior knowledge, we can draw a hypothetical causal diagram in Figure \ref{fig:causal_apply}.

\begin{figure}
\centering
\includegraphics[scale=0.6]{figure/casual_diagram_apply_p2.png}
\caption{A hypothetical causal diagram for application study. The black arrows refer to the causal relationship among confounders, the exposure, and the outcome. The double-sided dash arrows refer to the associational relationship among the missing indicator, the outcome, and the covariates. The red arrow is the effect of primary interest. The dashed short line refers to no association between the missing indicator and the exposure given covariates and the outcome due to MAR assumption.} 
\label{fig:causal_apply}
\end{figure}

For simplicity, we rename the age variable as $X_1$, diabetes as $X_2$, sex as $X_3$, CVD as the exposure variable $A$, and death as the outcome variable $Y$. Then, following the same model setup in the simulation studies described in Section \ref{sec:sim}, we consider five different imputations and four different PS models as candidate models. 


\begin{table}
\centering
\caption{Results of model selection for COVID-19 data using IPW and DR methods} 
\label{table:apply_cor}
\resizebox{1\textwidth}{!}{
\begin{tabular}{lcccccccc}
\hline
Imp Model & PS Model & IPW Estimate  & DR Estimate  & $\text{Accuracy}^{(w)}$ & Out BIC & ASMD & KS & Rank Score \\ 
\hline
$A\sim X_1+X_2$ & $A^{\text{imp}} \sim X_1$ & 0.945 & 0.986 & 0.607 & 3790.876 & 0.050 & 0.292 & 7.5 \\ 
  $A\sim X_1+X_2$ & $A^{\text{imp}} \sim X_1+X_2$& 0.961 & 0.988 & 0.607 & 3796.287 & 0.006 & 0.290 & 12.0 \\ 
  $A\sim X_1+X_2$ & $A^{\text{imp}} \sim X_1+X_3$& 0.950 & 0.991 & 0.607 & 3789.885 & 0.048 & 0.292 & 5.0 \\ 
  $A\sim X_1+X_2$& $A^{\text{imp}} \sim X_1+X_2+X_3$ & 0.963 & 0.989 & 0.607 & 3795.815 & 0.008 & 0.290 & 9.5 \\ 
  \hdashline
  $A\sim X_1+X_3$ & $A^{\text{imp}} \sim X_1$  & 0.946 & 0.983 & 0.581 & 3790.743 & 0.042 & 0.267 & 12.0 \\ 
  $A\sim X_1+X_3$ &$A^{\text{imp}} \sim X_1+X_2$  & 0.960 & 0.984 & 0.581 & 3796.138 & 0.006 & 0.264 & 16.5 \\ 
  $A\sim X_1+X_3$ &$A^{\text{imp}} \sim X_1+X_3$ & 0.951 & 0.988 & 0.581 & 3789.763 & 0.040 & 0.267 & 9.5 \\ 
  $A\sim X_1+X_3$ &$A^{\text{imp}} \sim X_1+X_2+X_3$  & 0.962 & 0.986 & 0.581 & 3795.679 & 0.007 & 0.264 & 14.5 \\ 
    \hdashline
 $A\sim X_1+X_2+X_3$ & $A^{\text{imp}} \sim X_1$ & 0.942 & 0.984 & 0.606 & 3790.866 & 0.050 & 0.294 & 11.0 \\ 
  $A\sim X_1+X_2+X_3$ &$A^{\text{imp}} \sim X_1+X_2$ & 0.958 & 0.985 & 0.606 & 3796.309 & 0.006 & 0.291 & 16.5 \\  
  $A\sim X_1+X_2+X_3$ &$A^{\text{imp}} \sim X_1+X_3$ & 0.947 & 0.990 & 0.606 & 3789.882 & 0.049 & 0.294 & 8.5 \\ 
  $A\sim X_1+X_2+X_3$ &$A^{\text{imp}} \sim X_1+X_2+X_3$  & 0.960 & 0.986 & 0.606 & 3795.837 & 0.008 & 0.291 & 14.0 \\ 
    
    \hdashline
  $A\sim X_1+X_2+Y$ &$A^{\text{imp}} \sim X_1$ & 0.942 & 0.985 & 0.606 & 3790.846 & 0.050 & 0.293 & 8.5 \\ 
  $A\sim X_1+X_2+Y$ &$A^{\text{imp}} \sim X_1+X_2$ & 0.959 & 0.986 & 0.606 & 3796.272 & 0.006 & 0.291 & 13.5 \\ 
  $A\sim X_1+X_2+Y$ &$A^{\text{imp}} \sim X_1+X_3$ & 0.947 & 0.990 & 0.606 & 3789.852 & 0.049 & 0.293 & 6.0 \\ 
  $A\sim X_1+X_2+Y$ &$A^{\text{imp}} \sim X_1+X_2+X_3$  & 0.961 & 0.987 & 0.606 & 3795.797 & 0.008 & 0.291 & 11.0 \\ 
    \hdashline
  $A\sim X_1+X_2+X_3+Y$ & $A^{\text{imp}} \sim X_1$ & 0.946 & 0.986 & 0.608 & 3790.742 & 0.048 & 0.289 & 3.5 \\ 
  $A\sim X_1+X_2+X_3+Y$ & $A^{\text{imp}} \sim X_1+X_2$ & 0.960 & 0.987 & 0.608 & 3796.144 & 0.006 & 0.287 & 9.0 \\ 
  $A\sim X_1+X_2+X_3+Y$ &$A^{\text{imp}} \sim X_1+X_3$ & 0.950 & 0.991 & 0.608 & 3789.735 & 0.048 & 0.289 & \textbf{1.0} \\ 
  $A\sim X_1+X_2+X_3+Y$ &$A^{\text{imp}} \sim X_1+X_2+X_3$ & 0.962 & 0.988 & 0.608 & 3795.661 & 0.008 & 0.287 & 6.0 \\ 
   \hline
\end{tabular}
}

\end{table}


Without assuming the hypothetical DAG is correct,  we apply the rank score to select the best imputation and PS models among the candidates. Table \ref{table:apply_cor} reports the estimated causal effects and the values of each criteria. Based on the proposed rank score, the best imputation model includes ``age'', ``sex'', ``diabetes'', and ``death'' as the predictors, and the best PS model selects ``age'' and ``sex'' as the predictors. This is consistent with the simulation results, i.e., the proposed criterion can still help us select the full imputation model plus the outcome-related PS model as the best choice following the hypothetical DAG we draw in Figure~\ref{fig:causal_apply}.  


\subsection{Estimation Results}
\label{sec:apply_est}

After the model selection, we obtain the estimated causal effect of CVD on mortality using IPW and DR estimators. We also provide confidence intervals (CI) after running $B=2000$ bootstrap resamples, as shown in Table \ref{table:apply_est}. We compare two different types of CI including a CI based on bootstrap standard error (BSE) with a normal approximation, called ``CI Normal'', and another CI based on bootstrap percentiles (between 2.5\% and 97.5\% percentiles of bootstrap point estimates), called ``CI Percentile'' \cite{tibshirani1993introduction}. 

The estimated causal risk ratio of CVD on the mortality for COVID-19 patients is 0.95 using IPW method with a \text{95\% CI of [0.867, 1.052] using CI Percentile}. For DR method, we also obtain a similar result as IPW method. 


Although some researchers may presume that patients with CVD status increase the risk of death, our analysis does not support this assumption. Since CI covers the value of 1, we would expect that the statistical test for whether the ratio of patients with CVD is higher than those without CVD is not significant after adjusting for missing data and confounding factors. One possible explanation for this finding is due to the confounding issue, i.e. age is strongly associated with both CVD status and death. Additionally, researchers may have ignored the effect of missing exposure on the estimated results. These results align with recent clinical research \cite{di,vasbinder2022relationship}, which suggests that those CVD risk factors, such as age, rather than CVD status itself, are the primary contributors to mortality.

One of the limitations of the application study is that we only include four clinical variables. Since we lack enough observations of other variables in the real dataset, we may miss some relevant confounders in the analysis. In addition, the model selection and estimated results are only based on the samples in Brazil instead of the world population.

\begin{table}
\centering
\caption{Estimated results using IPW and DR estimators for COVID-19 data} 
\label{table:apply_est}
\begin{tabular}{lccccc}
  \hline
Method & Selected Imputation, PS Models & Estimate (risk ratio) & BSE & CI Normal & CI Percentile \\ 
  \hline
IPW &  $ A\sim X_1+X_2+X_3+Y$, $A^{\text{imp}} \sim X_1+X_3$  & 0.950 & 0.048 & (0.856,1.044) & (0.867,1.052)  \\  
  DR &  $ A\sim X_1+X_2+X_3+Y$, $A^{\text{imp}} \sim X_1+X_3$  & 0.991 & 0.050 & (0.897,1.085) & (0.905,1.098) \\ 
   \hline
\end{tabular}
    \begin{tablenotes}
      \item  
    Estimate: causal effect (as the risk ratio) of CVD on mortality. BSE: bootstrap standard errors. CI-Normal: 95\% bootstrap normal confidence interval using BSE. CI-Percentile: 95\% bootstrap percentile confidence interval.
    \end{tablenotes}
\end{table}

\section{Discussion}

In causal inference, when the exposure is MAR, a common practice is to impute the missing values and then estimate causal effects using the imputed data. Model selection on both imputation and PS models is a challenging problem when the exposure is MAR, and simply selecting the imputation and PS models with the same set of exposure-related covariates is not appropriate. In addition, few studies discuss the appropriate criterion for selecting these models when the true value is unknown in the application studies. 

In this paper, we conduct simulation studies to investigate the effect of different imputation and PS models on the estimated causal effect. Unexpectedly, we find that even though the true missingness model does not depend on the outcome in our simulation setup, selecting the outcome and outcome-related variables into the imputation model can improve the accuracy of the imputed exposure and reduce the bias of the estimated causal effect. In addition,  we propose ``rank score'' as a rank-based and unit-free criterion to evaluate the performance of both imputation and PS models, which shows a higher correlation with RMSE in simulation studies. 

The limitation of the rank score is that it requires ranking both $\text{Accuracy}^{(w)}$ and BIC among all candidate models, which may be affected by the pool of candidate models. If the number of covariates in the dataset is small, we can recommend fitting all possible combinations of models.  On the other hand, when the number of candidate models is very large, ranking all candidate models becomes time-consuming. To address this issue, researchers can perform model selection sequentially, i.e. we use $\text{Accuracy}^{(w)}$ to select the best imputation model, and then we select PS model through BIC given that imputation model. However, such a sequential strategy is not ideal in some cases because the selection of PS model is only conditioned on the selected imputation model from the first step, which may not lead to the overall best models with the smallest RMSE. On the other hand, the rank score is similar to the ``best subset'' strategy because both accuracy and BIC contribute equally to the rank score, ensuring a balance between the rankings of imputation and PS models.

Further research can extend the proposed model selection criterion into multiple areas such as MAR on both exposure and outcome. Besides that, one can also consider other missing mechanisms such as MNAR. Additionally, it may be interesting for researchers to incorporate missing exposure into other complex settings, such as longitudinal or survival data in future studies.

\section{Acknowledgements}

We appreciate NSERC grants to support our research work and IntegraSUS \footnotemark \footnotetext{\url{https://integrasus.saude.ce.gov.br/}} for publicly sharing the COVID-19 data in Brazil. We would also like to thank Wei Liang for his great suggestions on the Appendix.

\bibliography{main}

\begin{thebibliography}{10}
\providecommand \doibase [0]{http://dx.doi.org/}%

\bibitem{rubin1976inference}
Rubin DB. Inference and missing data. {\it Biometrika} 1976\string; 63(3)\string: 581--592.

\bibitem{Bang}
Bang H, Robins JM. Doubly robust estimation in missing data and causal inference models. {\it Biometrics} 2005\string; 61(4)\string: 962--973.

\bibitem{williamson2012doubly}
Williamson EJ, Forbes A, Wolfe R. Doubly robust estimators of causal exposure effects with missing data in the outcome, exposure or a confounder. {\it Statistics in Medicine} 2012\string; 31(30)\string: 4382--4400.

\bibitem{zhang}
Zhang Z, Liu W, Zhang B, Tang L, Zhang J. Causal inference with missing exposure information: Methods and applications to an obstetric study. {\it Statistical Methods in Medical Research} 2016\string; 25(5)\string: 2053--2066.

\bibitem{rosenbaum1983central}
Rosenbaum PR, Rubin DB. The central role of the propensity score in observational studies for causal effects. {\it Biometrika} 1983\string; 70(1)\string: 41--55.

\bibitem{donders2006gentle}
Donders ART, Van Der~Heijden GJ, Stijnen T, Moons KG. A gentle introduction to imputation of missing values. {\it Journal of Clinical Epidemiology} 2006\string; 59(10)\string: 1087--1091.

\bibitem{Buuren}
Buuren vS, Boshuizen HC, Knook DL. Multiple imputation of missing blood pressure covariates in survival analysis. {\it Statistics in Medicine} 1999\string; 18(6)\string: 681--694.

\bibitem{buuren2011}
{van Buuren} S, Groothuis-Oudshoorn K. {mice}: Multivariate Imputation by Chained Equations in R. {\it Journal of Statistical Software} 2011\string; 45(3)\string: 1-67.
\newblock \href {\doibase 10.18637/jss.v045.i03} {doi: 10.18637/jss.v045.i03}

\bibitem{rubin2004multiple}
Rubin DB. {\it Multiple imputation for nonresponse in surveys}. 81.
\newblock John Wiley \& Sons .
\newblock 2004.

\bibitem{neyman1923application}
Neyman JS. On the application of probability theory to agricultural experiments. essay on principles. section 9.(tlanslated and edited by dm dabrowska and tp speed, statistical science (1990), 5, 465-480). {\it Annals of Agricultural Sciences} 1923\string; 10\string: 1--51.

\bibitem{scharfstein1999adjusting}
Scharfstein DO, Rotnitzky A, Robins JM. Adjusting for nonignorable drop-out using semiparametric nonresponse models. {\it Journal of the American Statistical Association} 1999\string; 94(448)\string: 1096--1120.

\bibitem{robins1994estimation}
Robins JM, Rotnitzky A, Zhao LP. Estimation of regression coefficients when some regressors are not always observed. {\it Journal of the American Statistical Association} 1994\string; 89(427)\string: 846--866.

\bibitem{kennedy2020efficient}
Kennedy EH. Efficient nonparametric causal inference with missing exposure information. {\it The International Journal of Biostatistics} 2020\string; 16(1)\string: 20190087.

\bibitem{Seaman}
Seaman SR, White IR. Review of inverse probability weighting for dealing with missing data. {\it Statistical Methods in Medical Research} 2013\string; 22(3)\string: 278--295.

\bibitem{brookhart2006variable}
Brookhart MA, Schneeweiss S, Rothman KJ, Glynn RJ, Avorn J, St{\"u}rmer T. Variable selection for propensity score models. {\it American Journal of Epidemiology} 2006\string; 163(12)\string: 1149--1156.

\bibitem{rubin2004principles}
Rubin DB. On principles for modeling propensity scores in medical research.. {\it Pharmacoepidemiology and Drug Safety} 2004\string; 13(12)\string: 855--857.

\bibitem{rosenbaum}
Rosenbaum PR, Rubin DB. The central role of the propensity score in observational studies for causal effects. {\it Biometrika} 1983\string; 70(1)\string: 41--55.

\bibitem{tanner2014identifying}
Tanner-Smith EE, Lipsey MW. Identifying baseline covariates for use in propensity scores: A novel approach illustrated for a nonrandomized study of recovery high schools. {\it Peabody Journal of Education} 2014\string; 89(2)\string: 183--196.

\bibitem{glymour2016causal}
Glymour M, Pearl J, Jewell NP. {\it Causal inference in statistics: A primer}.
\newblock John Wiley \& Sons .
\newblock 2016.

\bibitem{Thompson}
Horvitz DG, Thompson DJ. A generalization of sampling without replacement from a finite universe. {\it Journal of the American Statistical Association} 1952\string; 47(260)\string: 663--685.

\bibitem{lilliefors1967kolmogorov}
Lilliefors HW. On the Kolmogorov-Smirnov test for normality with mean and variance unknown. {\it Journal of the American Statistical Association} 1967\string; 62(318)\string: 399--402.

\bibitem{berger2014kolmogorov}
Berger VW, Zhou Y. Kolmogorov--smirnov test: Overview. {\it Wiley statsref: Statistics reference online} 2014.

\bibitem{franklin2014metrics}
Franklin JM, Rassen JA, Ackermann D, Bartels DB, Schneeweiss S. Metrics for covariate balance in cohort studies of causal effects. {\it Statistics in Medicine} 2014\string; 33(10)\string: 1685--1699.

\bibitem{bhattacharya2007instrumental}
Bhattacharya J, Vogt WB. Do instrumental variables belong in propensity scores?. {\it Working Paper 343, National Bureau of Economic Research} 2007.

\bibitem{zhu2015variable}
Zhu Y, Schonbach M, Coffman DL, Williams JS. Variable selection for propensity score estimation via balancing covariates. {\it Epidemiology} 2015\string; 26(2)\string: e14--e15.

\bibitem{westreich2011role}
Westreich D, Cole SR, Funk MJ, Brookhart MA, St{\"u}rmer T. The role of the c-statistic in variable selection for propensity score models. {\it Pharmacoepidemiology and Drug Safety} 2011\string; 20(3)\string: 317--320.

\bibitem{patrick2011implications}
Patrick AR, Schneeweiss S, Brookhart MA, et al. The implications of propensity score variable selection strategies in pharmacoepidemiology: an empirical illustration. {\it Pharmacoepidemiology and Drug Safety} 2011\string; 20(6)\string: 551--559.

\bibitem{kassir2020risk}
Kassir R. Risk of COVID-19 for patients with obesity. {\it Obesity Reviews} 2020\string; 21(6).

\bibitem{dana2020insight}
Dana PM, Sadoughi F, Hallajzadeh J, et al. An insight into the sex differences in COVID-19 patients: what are the possible causes?. {\it Prehospital and Disaster Medicine} 2020\string; 35(4)\string: 438--441.

\bibitem{abdi2020diabetes}
Abdi A, Jalilian M, Sarbarzeh PA, Vlaisavljevic Z. Diabetes and COVID-19: A systematic review on the current evidences. {\it Diabetes Research and Clinical Practice} 2020\string; 166\string: 108347.

\bibitem{bertoluci2017cardiovascular}
Bertoluci MC, Rocha VZ. Cardiovascular risk assessment in patients with diabetes. {\it Diabetology \& Metabolic Syndrome} 2017\string; 9\string: 1--13.

\bibitem{tibshirani1993introduction}
Tibshirani RJ, Efron B. An introduction to the bootstrap. {\it Monographs on Statistics and Applied Probability} 1993\string; 57\string: 1--436.

\bibitem{di}
Di~Castelnuovo A, Bonaccio M, Costanzo S, et al. Common cardiovascular risk factors and in-hospital mortality in 3,894 patients with COVID-19: survival analysis and machine learning-based findings from the multicentre Italian CORIST Study. {\it Nutrition, Metabolism and Cardiovascular Diseases} 2020\string; 30(11)\string: 1899--1913.

\bibitem{vasbinder2022relationship}
Vasbinder A, Meloche C, Azam TU, et al. Relationship between preexisting cardiovascular disease and death and cardiovascular outcomes in critically ill patients with COVID-19. {\it Circulation: Cardiovascular Quality and Outcomes} 2022\string; 15(10)\string: e008942.

\end{thebibliography}

\appendix

\renewcommand{\thesection}{A.\arabic{section}}
\renewcommand{\thesubsection}{A.\arabic{subsection}}
\renewcommand{\thetable}{A.\arabic{table}}
\renewcommand{\thefigure}{A.\arabic{figure}}
\renewcommand{\theequation}{A.\arabic{equation}}

\subsection{Proof of the Validity of Weighted Accuracy}
\label{app:proof_acc}

The key idea of weighted accuracy is to create a pseudo sample after weighting on those observed individuals so that the weighted accuracy on the observed dataset can represent the original accuracy on the whole dataset. Based on MAR assumption, if the missingness model is correct, under the true value of $w_i$, we can prove that,
\begin{equation}
\label{proof:ipw_acc}
\begin{aligned}
\text{$\text{Accuracy}^{(w)}$}(A^{\text{imp}}) \xrightarrow[]{\text{p}} 
&E \left[\frac{1-R_i}{1-w_i} \mathbbm{1}(A_i=A^{\text{imp}}) \right] \frac{E(\delta_i)}{E(\delta_i)}, \quad \mbox{as} \quad n \rightarrow \infty\\
&=E\left\{ \frac{E[1-R_i|\bm{X},Y]}{1-w_i}  E[\mathbbm{1}(A_i=A^{\text{imp}})|\bm{X},Y] \right\} \\
&=E\left\{  E[\mathbbm{1}(A_i=A^{\text{imp}})|\bm{X},Y] \right\}
=P(A_i=A^{\text{imp}}).
\end{aligned}
\end{equation}
Here the first equality holds because $\delta_i$ is an indicator of whether the individual is randomly selected into the testing data or not, which is free of $(\bm{X},Y)$. Its expectation is an arbitrarily chosen ratio, which can also be cancelled. The second equality holds because after taking double expectation with the condition of $(\bm{X},Y)$, we know $A_i \bot R_i|\bm{X},Y$ based on the MAR assumption. The third equality holds because the full missingness model is correctly specified, so we know $E(R_i|\bm{X},Y)=w_i$. After taking the double expectation again, we know $\text{Accuracy}^{(w)}$ will converge to the true accuracy on the original data as $n \to \infty$. 

Notice that Equation \ref{proof:ipw_acc} also describes why we cannot only use the naive approach to estimate the accuracy on the observed data, written as $\text{Naive-Accuracy}=\sum_{i=1}^n (1-R_i)\mathbbm{1}(A_i=A^{\text{imp}})$. In such a way, the missing indicator will largely affect the naive estimator, which cannot be cancelled without the weights $w_i$. Therefore, to properly evaluate the performance of the imputation model, we need to adopt the inverse weights of the missingness on the estimated accuracy of the observed data.

In the simulation studies, since we know the true missing values, we can also impute all missing values on the whole data via MICE and calculate the standard accuracy, called the ``Benchmark-Accuracy'' on the original data. In such a way, the bias between $\text{Accuracy}^{(w)}$ and Benchmark-Accuracy is close to 0.008 (the results are omitted in the simulation section), which supports the statement that the weighted accuracy will converge to the true accuracy as $n \to \infty$.  However, we will not know the true accuracy in the application due to the unknown missing data, so $\text{Accuracy}^{(w)}$ will be a useful tool for researchers to evaluate the selected imputation model.



\subsection{Estimate Risk Ratio for Binary Outcome using DR method}
\label{sec_simdr_binary}

\begin{table}[h]
\centering
\caption{Performance of different models using DR method when $n=500$} 
\label{table:sim_dr_n500}
\begin{tabular}{lcccccccccc}
\hline
Imp, PS Models & Bias & Bias Rate  & ESE& RMSE \\ 
\hline
Exp, naive & -0.186 & -12.198 & 0.084 & 0.204 \\ 
  Exp, exp & -0.222 & -14.553 & 0.085 & 0.237 \\ 
  Exp, out & -0.188 & -12.374 & 0.065 & 0.199 \\ 
  Exp, covs & -0.225 & -14.804 & 0.066 & 0.235 \\ 
    \hdashline
  Out, naive & -0.291 & -19.087 & 0.109 & 0.310 \\ 
  Out, exp & -0.321 & -21.056 & 0.103 & 0.337 \\ 
  Out, out & -0.226 & -14.859 & 0.058 & 0.233 \\ 
  Out, covs & -0.264 & -17.354 & 0.057 & 0.270 \\ 
    \hdashline
  Covs, naive & -0.182 & -11.933 & 0.123 & 0.219 \\ 
  Covs, exp & -0.215 & -14.105 & 0.134 & 0.253 \\ 
  Covs, out & -0.189 & -12.392 & 0.065 & 0.200 \\ 
  Covs, covs & -0.225 & -14.786 & 0.067 & 0.235 \\ 
    \hdashline
  Res, naive & 0.102 & 6.679 & 0.176 & 0.204 \\ 
  Res, exp & 0.133 & 8.763 & 0.211 & 0.250 \\ 
  Res, out & -0.056 & -3.654 & 0.100 & 0.114 \\ 
  Res, covs & -0.066 & -4.350 & 0.115 & 0.132 \\ 
    \hdashline
  Full, naive & -0.017 & -1.102 & 0.160 & 0.161 \\ 
  Full, exp & -0.015 & -0.977 & 0.189 & 0.189 \\ 
  Full, out & -0.010 & -0.660 & 0.104 & \textbf{0.104} \\ 
  Full, covs & -0.011 & -0.691 & 0.120 & 0.120 \\ 
   \hline
\end{tabular}
 \begin{tablenotes}
      \item  
 Here, Exp, Out, Res, and Covs, refer to the exposure-related, outcome-related, response-included, and covariates-included models, respectively. The lowest RMSE is bolded above. 
    \end{tablenotes}
\end{table}

\begin{table}[h]
\centering
\caption{Criterion values for different models  using DR method when $n=500$} 
\label{table:sim_dr_n500_acc}
\begin{tabular}{lcccccccccc}
 \hline
Imp, PS Models & $\text{Accuracy}^{(w)}$ & Out BIC & ASMD & KS & ABIC& Rank Score & Rank(RMSE) \\ 
 \hline
Exp, naive & 0.586 & 677.561 & 0.161 & 0.130 &  0.608 & 11.789 & 10 \\ 
  Exp, exp & 0.586 & 720.437 & 0.022 & 0.130 & 0.675 & 14.272 & 15 \\ 
  Exp, out & 0.586 & 430.074 & 0.149 & 0.130 & 0.225 & 7.849 & 7 \\ 
  Exp, covs & 0.586 & 433.966 & 0.014 & 0.130 & 0.231 & 8.959 & 14 \\ 
    \hdashline
  Out, naive & 0.507 & 683.862 & 0.103 & 0.148 & 0.930 & 15.947 & 19 \\ 
  Out, exp & 0.507 & 724.226 & 0.031 & 0.148 & 0.993 & 18.306 & 20 \\ 
  Out, out & 0.507 & 434.737 & 0.075 & 0.148 & 0.545 & 12.460 & 12 \\ 
  Out, covs & 0.507 & 435.659 & 0.008 & 0.148 &  0.546 & 12.918 & 18 \\ 
    \hdashline
  Covs, naive & 0.585 & 676.668 & 0.170 & 0.136 & 0.612 & 11.875 & 11 \\ 
  Covs, exp & 0.585 & 719.239 & 0.033 & 0.136 & 0.678 & 14.354 & 17 \\ 
  Covs, out & 0.585 & 430.103 & 0.148 & 0.136 & 0.231 & 8.043 & 8 \\ 
  Covs, covs & 0.585 & 433.958 & 0.014 & 0.136 & 0.237 & 9.136 & 13 \\ 
    \hdashline
  Res, naive & 0.609 & 652.471 & 0.182 & 0.124 & 0.484 & 8.103 & 9 \\ 
  Res, exp & 0.609 & 694.546 & 0.040 & 0.124 & 0.549 & 10.450 & 16 \\ 
  Res, out & 0.609 & 411.095 & 0.154 & 0.124 & 0.110 & 3.993 & 2 \\ 
  Res, covs & 0.609 & 415.906 & 0.015 & 0.124 & 0.118 & 4.580 & 4 \\ 
    \hdashline
  Full, naive & 0.622 & 662.369 & 0.172 & 0.138 & 0.430 & 7.463 & 5 \\ 
  Full, exp & 0.622 & 705.051 & 0.034 & 0.138 & 0.497 & 9.941 & 6 \\ 
  Full, out & 0.622 & 401.525 & 0.149 & 0.138 & 0.027 & \textbf{1.989} & \textbf{1} \\ 
  Full, covs & 0.622 & 406.367 & 0.014 & 0.138 &  0.034 & 2.574 & 3 \\ 
   \hline
\end{tabular}
 \begin{tablenotes}
      \item  
$\text{Accuracy}^{(w)}$: weighted accuracy on the observed data; Out BIC: BIC of the outcome model; ASMD: absolute standardized mean difference for $(X_1,X_2,X_3)$; KS: Kolmogorov-Smirnov distance for $(X_1,X_2,X_3)$; ABIC: rescaled product of Accuracy and BIC. Rank Score: average ranks of $1-\text{Accuracy}^{(w)}$ and BIC. The lowest rank and RMSE are bolded above. 
    \end{tablenotes}
\end{table}

\begin{sidewaysfigure}
\centering
\includegraphics[width=1\textwidth,height=10cm]{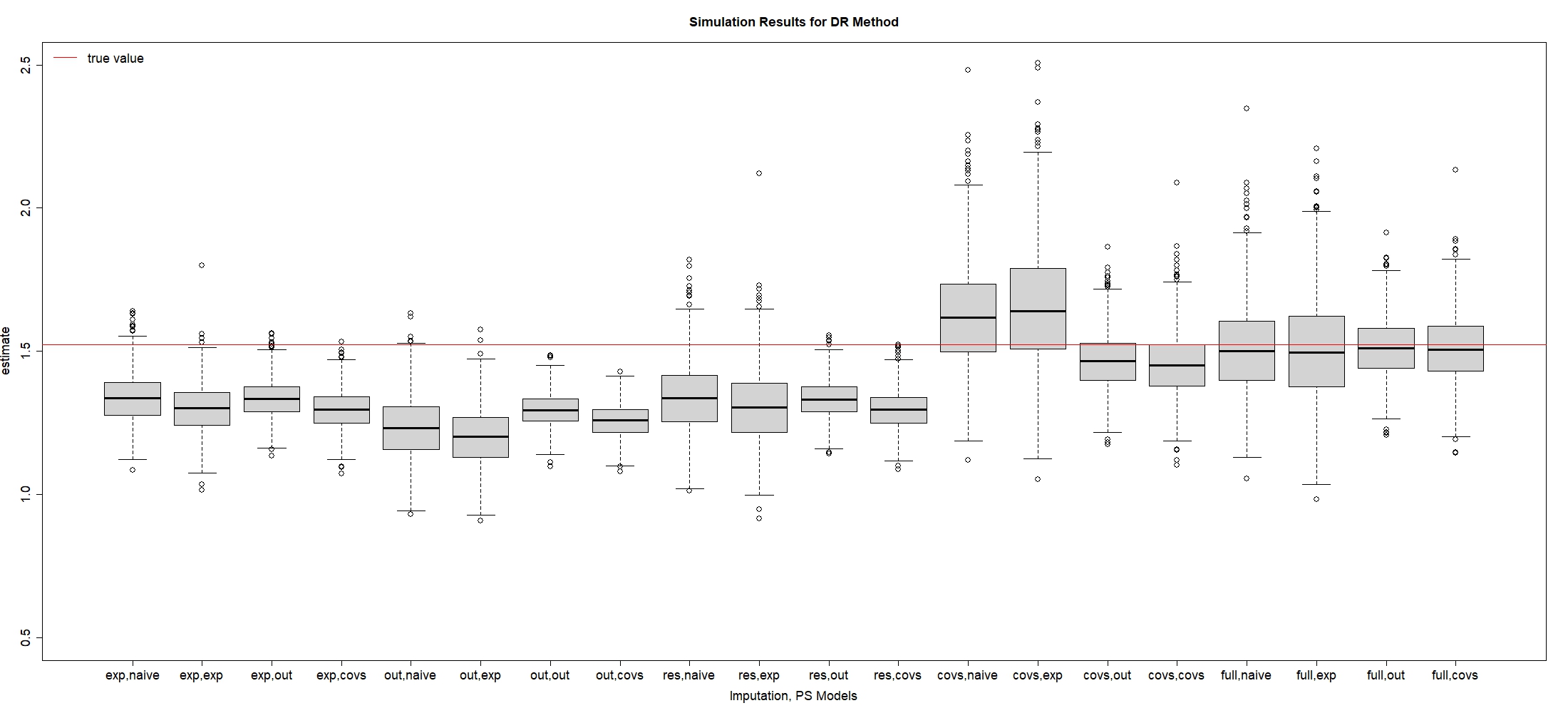}
\caption{Boxplot of estimated values for DR estimators in the simulation studies when $n=500$. The x-axis refers to the combination of imputation and PS models. Here, Exp, Out, Res, and Covs, refer to the exposure-related, outcome-related, response-included, and covariates-included models, respectively. The red line is the true causal effect. } 
\label{figure:sim_dr_n500}
\end{sidewaysfigure}

\end{document}


\title{Supplementary Material for  ``Model Selection for Causal Modeling in Missing Exposure Problems"}

\author[1]{Yuliang Shi*}

\author[1]{Yeying Zhu}

\author[1]{Joel A. Dubin}

\authormark{Yuliang \textsc{et. al.}}

\maketitle


\section{Estimation of ACE for a Continuous Outcome}
\label{sec_sim_cont}

In this section, we provide the selection and estimation results when we intend to estimate the average causal effect (ACE), defined as $\tau=\tau_1-\tau_0$ for a continuous outcome $Y$. Except for the changes in the outcome model, the setup for other models are the same as before. More specifically,  $Y, A$, and the missing indicator $R$ are generated by the following three models:

\begin{itemize}
    \item missingness model: $\text{logit}\{P(R=1|\bm{X})\}=-0.3+0.4X_1+0.6X_2+1.8X_3$;
    \item treatment model: $\text{logit}\{P(A=1|\bm{X})\}=-0.2+0.3X_1+1X_2$;
    \item outcome model: $E(Y|A,\bm{X})=-0.2+2A-0.3X_1+2.5X_3;$
\end{itemize}

\begin{figure}[h]
\centering
\includegraphics[width=\textwidth,height=10cm]{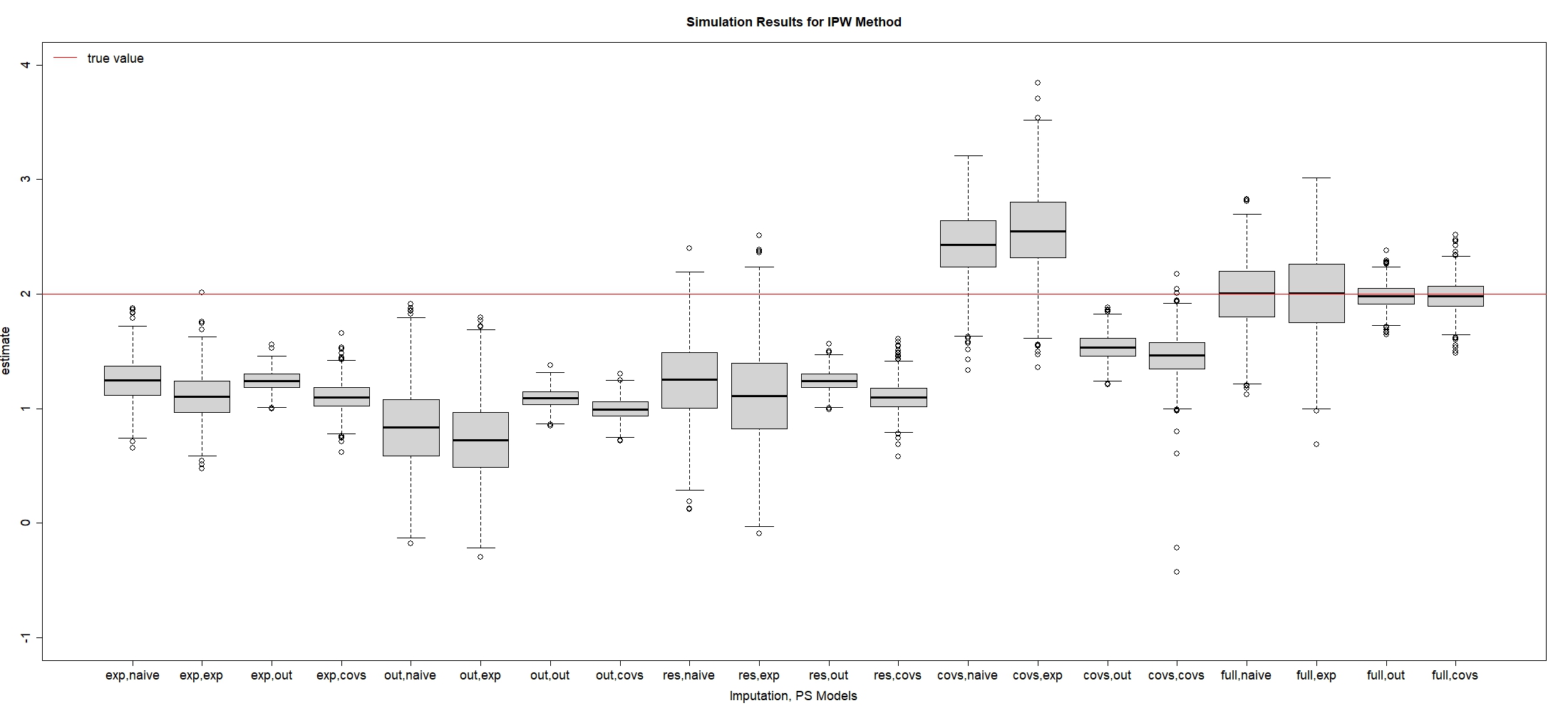}
\caption{Boxplot of estimated values for IPW estimators in the simulation studies for the continuous outcome to estimate ACE when $n=500$. The x-axis refers to which imputation and PS models are used. Here, Exp, Out, Res, and Covs, refer to the exposure-related, outcome-related, response-included, and covariates-included models, respectively. The red line is a true causal effect. } 
\label{figure:sim_ipw_n500_lm_ace}
\end{figure}

\begin{figure}[h]
\centering
\includegraphics[width=1\textwidth,height=10cm]{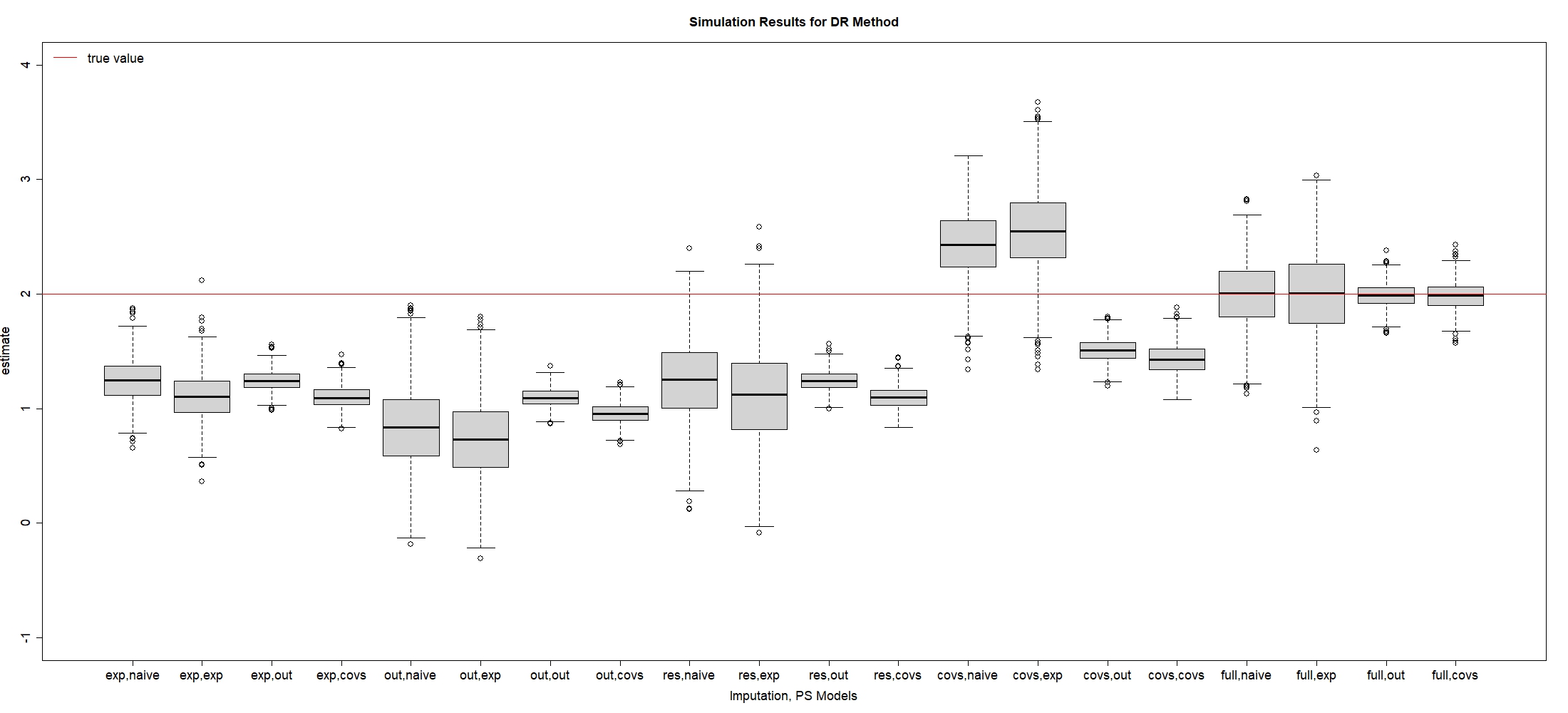}
\caption{Boxplot of estimated values for DR estimators in the simulation studies for the continuous outcome to estimate ACE when $n=500$. The x-axis refers to which imputation and PS models are used. Here, Exp, Out, Res, and Covs, refer to the exposure-related, outcome-related, response-included, and covariates-included models, respectively. The red line is the true causal effect. } 
\label{figure:sim_dr_n500_lm_ace}
\end{figure}

The true causal effect $\tau=2$ in this setup. Tables \ref{table:sim_ipw_n500_lm_ace} and \ref{table:sim_dr_n500_lm_ace} display the simulation results for estimating ACE using IPW and DR methods, respectively, while Figures \ref{figure:sim_ipw_n500_lm_ace} and \ref{figure:sim_ipw_n500_lm_ace} visualize the results. Similar to the binary outcome case, \ref{figure:sim_ipw_n500_lm_ace} and \ref{figure:sim_dr_n500_lm_ace} show that the full imputation model plus the outcome-related PS model leads to the smallest RMSE among other candidate models. In contrast, the outcome-related imputation model plus the exposure-related PS model leads to the largest RMSE, which is the worst model. Additionally, Tables~\ref{table:sim_ipw_n500_lm_acc} and \ref{table:sim_dr_n500_lm_acc} show the value of different criteria for each combination of the imputaion and PS models while Table~\ref{table:sim_cor_n500_lm_ace} displays the correlation of different model selection criteria with RMSE. The correlation of the rank score with RMSE is higher than 0.8 and the rank score is the smallest value for the full imputation plus the outcome-related PS model. Even though the correlation between weighted accuracy and RMSE is also higher than 0.8 in this case, we cannot select the best PS model if we only use weighted accuracy. Compared with other traditional criteria, the rank score combines the performance of both models, so it can successfully pick up the best models together. 

\begin{table}[h]
\caption{Performance of different models using IPW method when $n=500$ for continuous outcome to estimate ACE} 
\label{table:sim_ipw_n500_lm_ace}
\centering
\begin{tabular}{lcccc}
 \hline
Imp, PS Models & Bias & Bias Rate & ESE  & RMSE \\ 
   \hline
Exp, naive & -0.754 & -37.708 & 0.186 & 0.777 \\ 
  Exp, exp & -0.895 & -44.766 & 0.203 & 0.918 \\ 
  Exp, out & -0.757 & -37.865 & 0.088 & 0.762 \\ 
  Exp, covs & -0.898 & -44.885 & 0.122 & 0.906 \\ 
  \hdashline
  Out, naive & -1.163 & -58.158 & 0.366 & 1.219 \\ 
  Out, exp & -1.277 & -63.846 & 0.366 & 1.328 \\ 
  Out, out & -0.910 & -45.479 & 0.082 & 0.913 \\ 
  Out, covs & -1.005 & -50.270 & 0.090 & 1.009 \\ 
    \hdashline
  Covs, naive & -0.752 & -37.589 & 0.358 & 0.833 \\ 
  Covs, exp & -0.889 & -44.452 & 0.419 & 0.983 \\ 
  Covs, out & -0.758 & -37.915 & 0.088 & 0.763 \\ 
  Covs, covs & -0.897 & -44.874 & 0.123 & 0.906 \\ 
    \hdashline
  Res, naive & 0.428 & 21.390 & 0.299 & 0.522 \\ 
  Res, exp & 0.548 & 27.392 & 0.362 & 0.656 \\ 
  Res, out & -0.466 & -23.292 & 0.113 & 0.479 \\ 
  Res, covs & -0.538 & -26.893 & 0.195 & 0.572 \\ 
    \hdashline
  Full, naive & -0.001 & -0.033 & 0.298 & 0.298 \\ 
  Full, exp & 0.000 & 0.012 & 0.361 & 0.361 \\ 
  Full, out & -0.018 & -0.922 & 0.101 & 0.103 \\ 
  Full, covs & -0.019 & -0.935 & 0.141 & 0.143 \\ 
   \hline
\end{tabular}
 \begin{tablenotes}
      \item  
  Here, Exp, Out, Res, and Covs, refer to the exposure-related, outcome-related, response-included, and covariates-included models, respectively. 
    \end{tablenotes}
\end{table}

\begin{table}[h]
\centering
\caption{Performance of different models using DR method to estimate ACE for continuous outcome when $n=500$ } 
\label{table:sim_dr_n500_lm_ace}
\begin{tabular}{lcccccccccc}
\hline
Imp, PS Models & Bias & Bias Rate  & ESE& RMSE \\ 
  \hline
Exp, naive & -0.754 & -37.718 & 0.186 & 0.777 \\ 
  Exp, exp & -0.897 & -44.845 & 0.204 & 0.920 \\ 
  Exp, out & -0.757 & -37.831 & 0.088 & 0.762 \\ 
  Exp, covs & -0.901 & -45.051 & 0.097 & 0.906 \\ 
    \hdashline
  Out, naive & -1.163 & -58.174 & 0.366 & 1.220 \\ 
  Out, exp & -1.271 & -63.529 & 0.367 & 1.323 \\ 
  Out, out & -0.903 & -45.130 & 0.082 & 0.906 \\ 
  Out, covs & -1.043 & -52.127 & 0.087 & 1.046 \\ 
    \hdashline
  Covs, naive & -0.752 & -37.599 & 0.358 & 0.833 \\ 
  Covs, exp & -0.890 & -44.476 & 0.422 & 0.984 \\ 
  Covs, out & -0.756 & -37.817 & 0.088 & 0.761 \\ 
  Covs, covs & -0.901 & -45.043 & 0.098 & 0.906 \\ 
    \hdashline
  Res, naive & 0.428 & 21.389 & 0.299 & 0.522 \\ 
  Res, exp & 0.551 & 27.542 & 0.363 & 0.660 \\ 
  Res, out & -0.492 & -24.621 & 0.103 & 0.503 \\ 
  Res, covs & -0.569 & -28.453 & 0.133 & 0.584 \\ 
    \hdashline
  Full, naive & -0.001 & -0.036 & 0.297 & 0.297 \\ 
  Full, exp & 0.003 & 0.130 & 0.363 & 0.363 \\ 
  Full, out & -0.018 & -0.878 & 0.101 & 0.102 \\ 
  Full, covs & -0.019 & -0.932 & 0.117 & 0.119 \\ 
   \hline
\end{tabular}
 \begin{tablenotes}
      \item  
  Here, Exp, Out, Res, and Covs, refer to the exposure-related, outcome-related, response-included, and covariates-included models, respectively. 
    \end{tablenotes}
\end{table}

\begin{table}[h]
\caption{Criterion values for different models using IPW method to estimate ACE for continuous outcome when $n=500$} 
\label{table:sim_ipw_n500_lm_acc}
\centering
\begin{tabular}{lcccccccc}
   \hline
Imp, PS Models & $\text{Accuracy}^{(w)}$ & Out BIC & ASMD & KS & ABIC & Rank Score & Rank(RMSE) \\ 
   \hline
Exp, naive & 0.592 & 2470.074 & 0.161 & 0.130 & 1.7032 & 13.364 & 11 \\ 
  Exp, exp & 0.592 & 2473.226 & 0.022 & 0.130 & 1.7062 & 13.946 & 16 \\ 
  Exp, out & 0.592 & 1680.469 & 0.149 & 0.130 & 0.9350 & 8.879 & 9 \\ 
  Exp, covs & 0.592 & 1676.146 & 0.014 & 0.130 & 0.9308 & 8.284 & 14 \\
    \hdashline
  Out, naive & 0.512 & 2481.170 & 0.102 & 0.148 & 1.9921 & 18.033 & 19 \\ 
  Out, exp & 0.512 & 2479.270 & 0.031 & 0.148 & 1.9903 & 17.950 & 20 \\ 
  Out, out & 0.512 & 1707.195 & 0.075 & 0.148 & 1.2392 & 13.368 & 15 \\ 
  Out, covs & 0.512 & 1684.670 & 0.008 & 0.148 & 1.2172 & 12.552 & 18 \\ 
    \hdashline
  Covs, naive & 0.591 & 2467.830 & 0.170 & 0.136 & 1.7068 & 13.376 & 12 \\ 
  Covs, exp & 0.591 & 2470.496 & 0.033 & 0.136 & 1.7094 & 13.835 & 17 \\ 
  Covs, out & 0.591 & 1680.727 & 0.148 & 0.136 & 0.9410 & 9.018 & 10 \\ 
  Covs, covs & 0.591 & 1676.198 & 0.014 & 0.136 & 0.9366 & 8.403 & 13 \\ 
    \hdashline
  Res, naive & 0.639 & 2396.528 & 0.219 & 0.153 & 1.4664 & 8.541 & 6 \\ 
  Res, exp & 0.639 & 2400.459 & 0.077 & 0.153 & 1.4703 & 8.949 & 8 \\ 
  Res, out & 0.639 & 1627.598 & 0.156 & 0.152 & 0.7184 & 4.648 & 5 \\ 
  Res, covs & 0.639 & 1628.720 & 0.018 & 0.152 & 0.7195 & 4.836 & 7 \\ 
    \hdashline
  Full, naive & 0.794 & 2429.053 & 0.167 & 0.130 & 0.9480 & 7.019 & 3 \\ 
  Full, exp & 0.794 & 2433.731 & 0.028 & 0.130 & 0.9526 & 7.489 & 4 \\ 
  Full, out & 0.794 & 1454.705 & 0.149 & 0.130 & 0.0004 & 1.024 & 1 \\ 
  Full, covs & 0.794 & 1459.247 & 0.014 & 0.130 & 0.0049 & 1.488 & 2 \\ 
   \hline
\end{tabular}
 \begin{tablenotes}
      \item  
   $\text{Accuracy}^{(w)}$: weighted accuracy on the observed data; Out BIC: BIC of the outcome model; ASMD: absolute standardized mean difference for $(X_1,X_2,X_3)$; KS: Kolmogorov-Smirnov distance for $(X_1,X_2,X_3)$; ABIC: rescaled product of Accuracy and BIC. Rank Score: average ranks of $(1-\text{Accuracy}^{(w)})$ and BIC. 
    \end{tablenotes}
\end{table}

\begin{table}[h]
\centering
\caption{Criterion values for different models  using DR method to estimate ACE when $n=500$} 
\label{table:sim_dr_n500_lm_acc}
\begin{tabular}{lcccccccccc}
 \hline
Imp, PS Models & $\text{Accuracy}^{(w)}$ & Out BIC & ASMD & KS & ABIC& Rank Score & Rank(RMSE) \\ 
\hline
Exp, naive & 0.592 & 2470.074 & 0.161 & 0.130 & 1.7032 & 13.364 & 11 \\ 
  Exp, exp & 0.592 & 2473.226 & 0.022 & 0.130 & 1.7062 & 13.946 & 16 \\ 
  Exp, out & 0.592 & 1680.469 & 0.149 & 0.130 & 0.9350 & 8.879 & 10 \\ 
  Exp, covs & 0.592 & 1676.146 & 0.014 & 0.130 & 0.9308 & 8.284 & 14 \\ 
    \hdashline
  Out, naive & 0.512 & 2481.170 & 0.102 & 0.148 & 1.9921 & 18.033 & 19 \\ 
  Out, exp & 0.512 & 2479.270 & 0.031 & 0.148 & 1.9903 & 17.950 & 20 \\ 
  Out, out & 0.512 & 1707.195 & 0.075 & 0.148 & 1.2392 & 13.368 & 15 \\ 
  Out, covs & 0.512 & 1684.670 & 0.008 & 0.148 & 1.2172 & 12.552 & 18 \\ 
    \hdashline
  Covs, naive & 0.591 & 2467.830 & 0.170 & 0.136 & 1.7068 & 13.376 & 12 \\ 
  Covs, exp & 0.591 & 2470.496 & 0.033 & 0.136 & 1.7094 & 13.835 & 17 \\ 
  Covs, out & 0.591 & 1680.727 & 0.148 & 0.136 & 0.9410 & 9.018 & 9 \\ 
  Covs, covs & 0.591 & 1676.198 & 0.014 & 0.136 & 0.9366 & 8.403 & 13 \\ 
    \hdashline
  Res, naive & 0.639 & 2396.528 & 0.219 & 0.153 & 1.4664 & 8.541 & 6 \\ 
  Res, exp & 0.639 & 2400.459 & 0.077 & 0.153 & 1.4703 & 8.949 & 8 \\ 
  Res, out & 0.639 & 1627.598 & 0.156 & 0.152 & 0.7184 & 4.648 & 5 \\ 
  Res, covs & 0.639 & 1628.720 & 0.018 & 0.152 & 0.7195 & 4.836 & 7 \\ 
    \hdashline
  Full, naive & 0.794 & 2429.053 & 0.167 & 0.130 & 0.9480 & 7.019 & 3 \\ 
  Full, exp & 0.794 & 2433.731 & 0.028 & 0.130 & 0.9526 & 7.489 & 4 \\ 
  Full, out & 0.794 & 1454.705 & 0.149 & 0.130 & 0.0004 & 1.024 & 1 \\ 
  Full, covs & 0.794 & 1459.247 & 0.014 & 0.130 & 0.0049 & 1.488 & 2 \\ 
   \hline
\end{tabular}
 \begin{tablenotes}
      \item  
$\text{Accuracy}^{(w)}$: weighted accuracy on the observed data; Out BIC: BIC of the outcome model; ASMD: absolute standardized mean difference for $(X_1,X_2,X_3)$; KS: Kolmogorov-Smirnov distance for $(X_1,X_2,X_3)$; ABIC: rescaled product of Accuracy and BIC. Rank Score: average ranks of $(1-\text{Accuracy}^{(w)})$ and BIC. 
    \end{tablenotes}
\end{table}

\begin{table}[h]
\centering
\caption{Spearman correlation between criteria and RMSE  for continuous outcome when $n=500$} 
\label{table:sim_cor_n500_lm_ace}
\begin{tabular}{lcc}
 \hline
Criterion & IPW Method & DR Method \\
  \hline
1-$\text{Accuracy}^{(w)}$ & 0.867 & 0.867 \\ 
  ASMD & -0.339 & -0.339 \\ 
  KS & 0.203 & 0.201 \\ 
  Out BIC & 0.583 & 0.583 \\ 
    \hdashline
  ABIC & 0.730 & 0.730 \\ 
  Rank Score & 0.854 & 0.854 \\ 
   \hline
\end{tabular}
\begin{tablenotes}
\item Here, ASMD: absolute standardized mean difference for $(X_1, X_2, X_3)$. KS: Kolmogorov-Smirnov distance for $(X_1,X_2,X_3)$. ABIC is the rescaled product of Accuracy and BIC. Rank Score: averaged ranks of $(1-\text{Accuracy}^{(w)})$ and BIC. 
\end{tablenotes}

\end{table}